\def\aas{{A\&AS}}
\def\actaa{{AcA}}

\def\esomess{{The Messenger}}
\def\nat{{Nature}}
\def\newa{{NewA}}

\def\cm2{cm$^{-2}$}

\def\c2{C~{\sc ii}}
\def\c4{C~{\sc iv}}
\def\fe2{Fe~{\sc ii}}
\def\fe3{Fe~{\sc iii}}
\def\mg1{Mg~{\sc i}}
\def\mg2{Mg~{\sc ii}}
\def\si2{Si~{\sc ii}}
\def\si4{Si~{\sc iv}}
\def\al2{Al~{\sc ii}}
\def\al3{Al~{\sc iii}}
\def\o1{O~{\sc i}}
\def\n1{N~{\sc i}}
\def\h1{H~{\sc i}}

\def\approxlt{\mathrel{\spose{\lower 3pt\hbox{$\sim$}}
        \raise 2.0pt\hbox{$<$}}}
\def\approxgt{\mathrel{\spose{\lower 3pt\hbox{$\sim$}}
        \raise 2.0pt\hbox{$>$}}}

\documentclass[article]{has40}
\usepackage{graphicx}
\usepackage{amssymb}
\tabletypesize{\normalsize}  

\def\plotone#1{\centering \leavevmode
\includegraphics[width=.95\columnwidth]{#1}}

\def\plottwo#1#2{\centering \leavevmode
\includegraphics[width=.45\columnwidth]{#1} \hfil
\includegraphics[width=.45\columnwidth]{#2}}

\def\plotone#1{\centering \leavevmode
\includegraphics[width=.95\columnwidth]{#1}}

\def\plottwo#1#2{\centering \leavevmode
\includegraphics[width=.45\columnwidth]{#1} \hfil
\includegraphics[width=.45\columnwidth]{#2}}

\usepackage{rotating}
\def\atel{{ATel}} 

\shortauthors{M. Catelan et al.}
\shorttitle{Stellar Variability in the VVV Survey}

\begin{document}
\large    
\pagenumbering{arabic}
\setcounter{page}{139}

\title{Stellar Variability in the VVV Survey:\\ Overview and First Results}

%
%
\author{{\noindent 
  M. Catelan,{$^{\rm 1,2}$} 
  D. Minniti,{$^{\rm 1,2}$}
  P. W. Lucas,{$^{\rm 3}$} 
  I. D\'ek\'any,{$^{\rm 1,2}$} 
  R. K. Saito,{$^{\rm 1,2,4}$} 
  R. Angeloni,{$^{\rm 1,2}$} J. Alonso-Garc\'{i}a,{$^{\rm 1,2}$} 
  M. Hempel,{$^{\rm 1,2}$} K. Helminiak,{$^{\rm 1,2,5}$} A. Jord\'an,{$^{\rm 1,2}$}
  R. Contreras Ramos,{$^{\rm 1,2}$} C. Navarrete,{$^{\rm 1,2}$} 
  J. C. Beam\'{i}n,{$^{\rm 1,2}$} A. F. Rojas,{$^{\rm 1,2}$} 
  F. Gran,{$^{\rm 1,2}$} C. E. Ferreira Lopes,{$^{\rm 1,2,6}$} 
  C. Contreras Pe\~na,{$^{\rm 3}$} 
  E. Kerins,{$^{\rm 7}$} L. Huckvale,{$^{\rm 7,8}$} M. Rejkuba,{$^{\rm 8}$}
  R. Cohen,{$^{\rm 9}$} F. Mauro,{$^{\rm 9}$} 
  J. Borissova,{$^{\rm 10}$} P. Amigo,{$^{\rm 1,2,10}$}
  S. Eyheramendy,{$^{\rm 11}$} K. Pichara,{$^{\rm 12}$}
  N. Espinoza,{$^{\rm 1,2}$} C. Navarro,{$^{\rm 1,2,10}$}
  G. Hajdu,{$^{\rm 1,2}$} D. N. Calder\'on Espinoza,{$^{\rm 1,2}$}
  G. A. Muro,{$^{\rm 1,2}$} H. Andrews,{$^{\rm 1,2,13}$}
  V. Motta,{$^{\rm 10}$} R. Kurtev,{$^{\rm 10}$}
  J. P. Emerson,{$^{\rm 14}$} C. Moni Bidin,{$^{\rm 2,15}$}
  A.-N. Chen\'e{$^{\rm 16}$}
\\
\\
{\it 
(1) Pontificia Universidad Cat\'olica de Chile, Instituto de Astrof\'{i}sica, Santiago, Chile\\
(2) The Milky Way Millennium Nucleus, Santiago, Chile\\
(3) University of Hertfordshire, Hatfield, UK \\
(4) Universidade Federal de Sergipe, S\~ao Crist\'ov\~ao, SE, Brazil\\
(5) Nicolaus Copernicus Astronomical Center, Toru\'{n}, Poland\\
(6) Universidade Federal do Rio Grande do Norte, Natal, Brazil\\
(7) The University of Manchester, Manchester, UK \\ 
(8) European Southern Observatory, Garching, Germany \\
(9) Universidad de Concepci\'on, Concepci\'on, Chile \\
(10) Universidad de Valpara\'{i}so, Valpara\'{i}so, Chile \\
(11) Pontificia Universidad Cat\'olica de Chile, Departamento de Estad\'{i}stica, Santiago, Chile\\
(12) Pontificia Universidad Cat\'olica de Chile, Facultad de Ingenier\'{i}a, Santiago, Chile\\
(13) Leiden Observatory, Leiden, The Netherlands\\ 
(14) Queen Mary, University of London, London, UK \\
(15) Instituto de Astronom\'{i}a, Universidad Cat\'olica del Norte, Antofagasta, Chile\\
(16) Gemini Observatory, Hawaii, USA\\
}
}
}

%
%


\begin{abstract}
The Vista Variables in the V\'{i}a L\'actea (VVV) ESO Public Survey is an ongoing time-series, 
near-infrared (IR) survey of the Galactic bulge and an adjacent portion of the inner disk, covering 562~square 
degrees of the sky, using ESO's VISTA telescope. The survey has provided superb multi-color photometry in 5 
broadband filters ($Z$, $Y$, $J$, $H$, and $K_s$), leading to the best map of the inner Milky Way ever 
obtained, particularly in the near-IR. The main variability part of the survey, which is focused on $K_s$-band observations, is 
currently underway, with bulge fields having been observed between 31 and 70 times, and 
disk fields between 17 and 36 times. When the survey is complete, bulge (disk) fields will 
have been observed up to a total of 100 (60) times, providing unprecedented depth and time coverage. 
Here we provide a first overview of stellar
variability in the VVV data, including examples of the light curves that have been collected thus
far, scientific applications, and our efforts towards the automated classification of VVV light 
curves. 
\end{abstract}

\section{The VVV Survey: A Brief Overview, and Current Status}\label{sec:intro}
\subsection{Overview}\label{sec:over}
\begin{sidewaysfigure}
    \centering
    \includegraphics[width=\textheight]{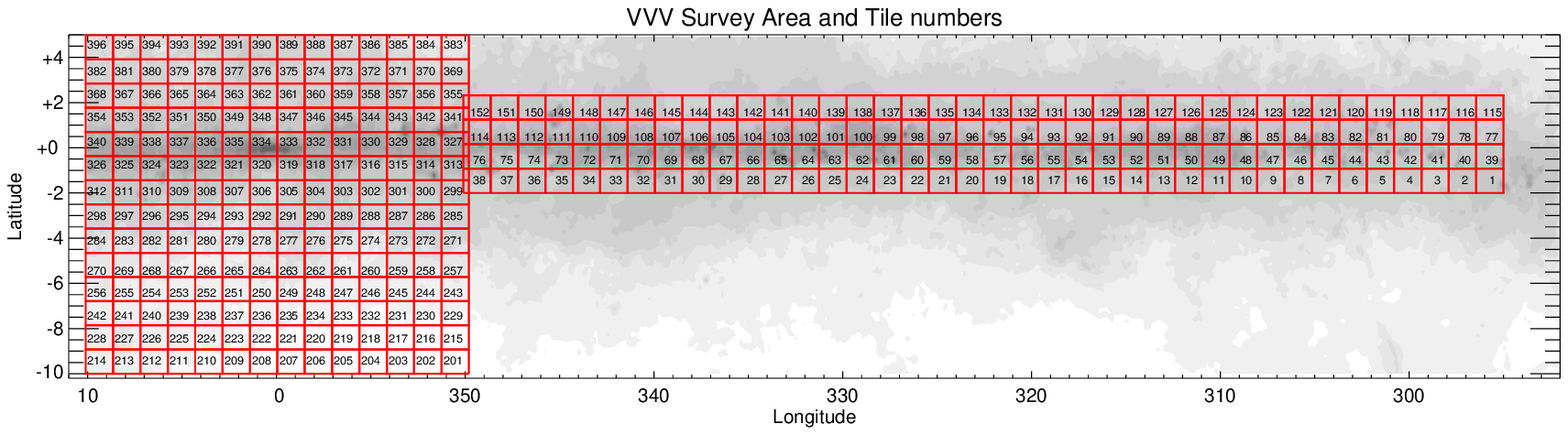}
\caption{VVV Survey area, superimposed on a gray-scale image that represents the Galactic dust extinction maps by \citep{schleg98}. Each of the 348 tiles of the survey is represented by a red rectangle, and is identified by its ID number. Each such tile, which covers a $\approx$~1.5 square degree field, has already been observed at least once in each of the $Z$, $Y$, $J$, $H$, and $K_s$ filters. By the end of the survey, each of these tiles will have been observed up to 100 times in $K_s$.}
\label{fig:area}
\end{sidewaysfigure}

The Vista Variables in the V\'{i}a L\'actea (VVV) ESO Public Survey \citep{dmea10,mcea11,rsea12} 
is a time-series, near-infrared (IR) survey of the Galactic bulge and an adjacent portion of the inner 
disk, covering 562~square degrees of the sky (Fig.~\ref{fig:area}). The survey has provided
multi-color photometry in 5 broadband filters ($Z$, $Y$, $J$, $H$, and $K_s$), but its main goal 
is to provide, for the first time, a homogeneous database 
for a variability study of the observed regions in the $K_s$-band. VVV has much improved photometric 
precision compared with, and extends much deeper than, 2MASS \citep{msea06}. In addition, in 
contrast to single-epoch surveys, which only allow the construction of 2-dimensional (2D) maps, 
with the addition of temporal information for well-established distance indicators such as RR Lyrae stars
\citep[e.g.,][]{lo86,dcarn95,gbea01,mcea04}, the VVV Survey will
enable us to resolve the 3D structure not only of the Milky Way, but also of the Sagittarius 
dwarf spheroidal galaxy \citep[Sgr dSph;][]{dalar96}, parts of which are also included in the VVV 
fields, and possibly even detect previously unknown Galactic structures and streams.


The survey is carried out at the Visible and Infrared Survey Telescope
for Astronomy (VISTA), a 4m-class telescope operated by ESO and
located at Cerro Paranal, Chile. The heart of the VISTA/VIRCAM
instrument is a $4\times 4$ array of Raytheon VIRGO IR detectors 
($2048\times 2048$ pixels), with a pixel size of $0\farcs34$ \citep{jeea06,gdea06}. 
The size of a uniformly covered field (also called a ``tile'') is 1.501~deg$^2$, hence the VVV
Survey requires a total of 348 such ``tiles'' to cover the survey area
and include a small overlap between neighboring tiles (Fig.~\ref{fig:area}).  

The data reduction is carried out at the Cambridge Astronomy Survey
Unit (CASU) in collaboration with the UK Wide-Field Astronomy
Unit (WFAU) in Edinburgh. Details about the data pipeline and the
various steps of the calibrations can be found in
\citet{demer04}, \citet{dhamb04,nhea08}, and \citet{dirw04}. Briefly, 
VVV images are pipeline-processed by the VISTA Data Flow System (VDFS; \citealt{demer04}), including all steps of data reduction from image processing to photometry and its calibration. Individual exposures are subjected to standard steps of pre-processing, such as flat-fielding, dark subtraction, and non-linearity correction. Science frames are composed of two dithered images, i.e. subsequent exposures taken with an offset of typically $\sim 40$ pixels, in order to  remove detector artifacts, cosmic rays, and other cosmetic defects. The resulting detector frame stacks, a.k.a. ``pawprints,'' have non-contiguous areal coverage due to the large gaps between the 16 VIRCAM chips. At each observational epoch, a sequence of six pawprints is acquired, and these are further combined to form a contiguous mosaic image, a.k.a. ``tile.'' 


\begin{figure*}
\centering \includegraphics[width=0.85\textwidth]{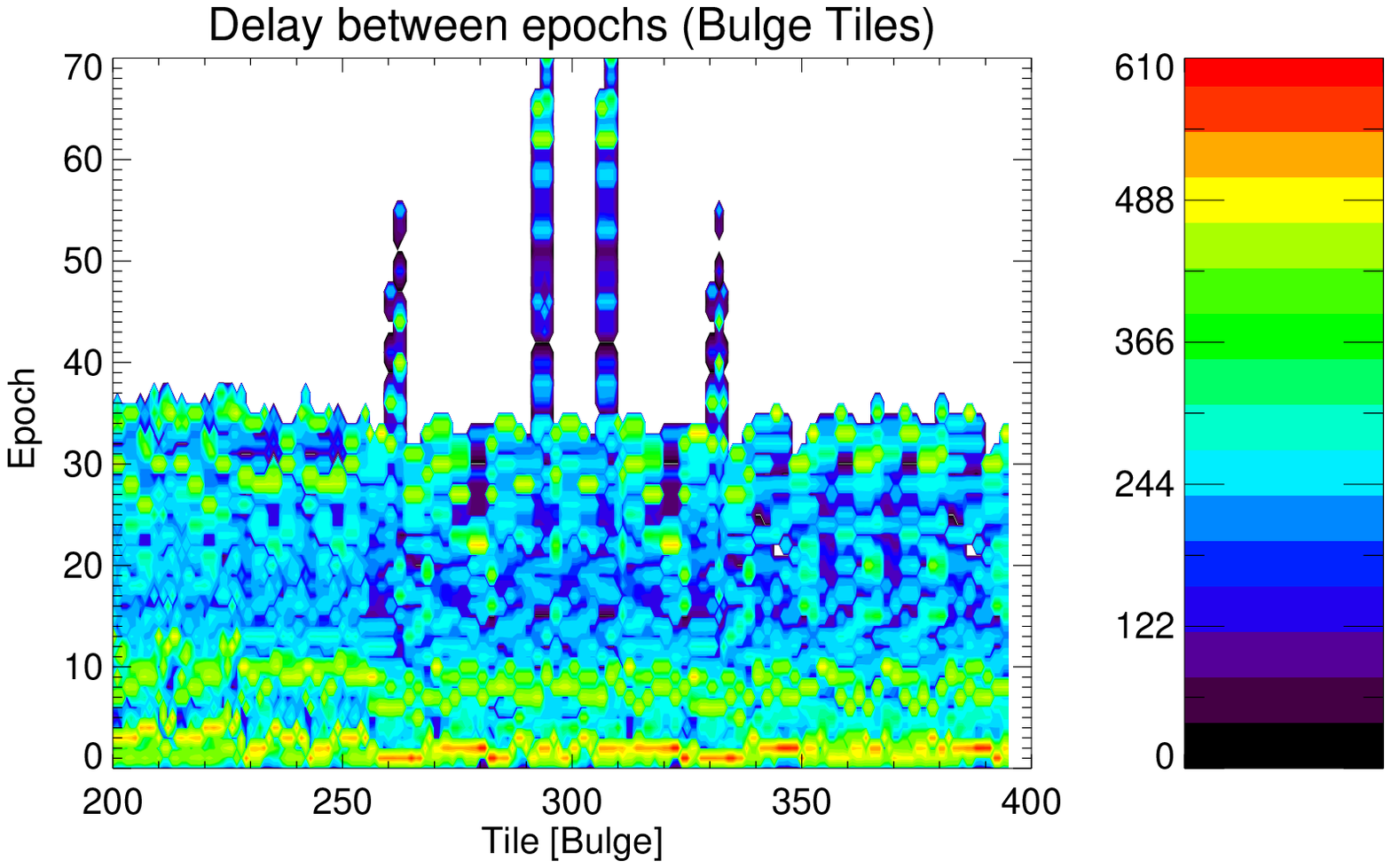}
\centering \includegraphics[width=0.85\textwidth]{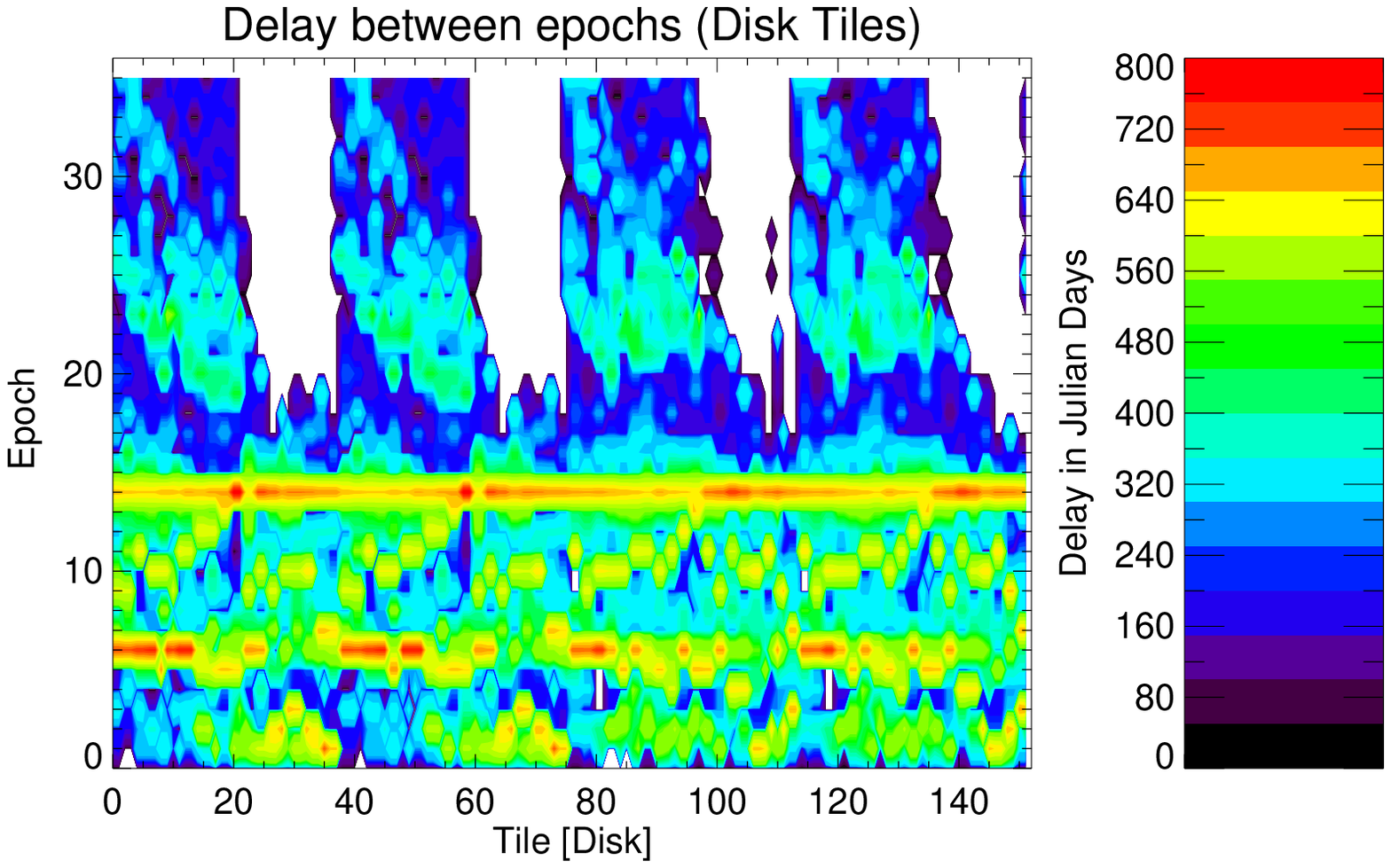}
\vskip0pt
\caption{Time delay (in Julian Days) between two consecutive epochs of $K_s$-band
  observations for the bulge ({\em top}) and disk ({\em bottom}) tiles. The 8 tiles with a large number of observing epochs (ID: 293, 294, 295, 296, 307, 308, 309, 310; see also Fig.~\ref{fig:area}) are bulge tiles situated in ``Baade's Window'' \citep{wb51}, with a much reduced  dust extinction and for which independent observations are available in the optical bands from the OGLE survey \citep{auea92}.}
\label{fig:epochs-delays}
\end{figure*}

Further steps in the reduction consist of source extraction and aperture photometry. Sets of small circular apertures with increasing radii are used in order to maximize the signal-to-noise ratio in highly time-varying seeing conditions, and to suppress systematics due to source crowding. Furthermore, the pipeline includes point-spread function (PSF) estimation and PSF fits for each object. Flux loss in the wings of the PSFs is remedied by aperture corrections \citep{dirw04}. Sources are classified based on the shape of the PSF \citep[see, e.g.,][]{rsea12}. Source positions are astrometrized using 2MASS \citep{msea06} stars as reference, with a median accuracy of $35-175$~mas, depending on magnitude \citep{rsea12}. Magnitudes are corrected for detector distortion, and in the case of the $JHK_s$ filters are zero-point (ZP) calibrated on a frame-by-frame basis using local 2MASS secondary standards. For the ZP calibration of the $Z$ and $Y$ filters, the procedure described in \citet{shea09} is followed. The ZP accuracy of $JHK_s$ magnitudes depends on the number of available non-saturated local standards (i.e., the sky conditions), and is usually within $1-2\%$, while the $ZY$ photometry is typically accurate to $\sim 0.05-0.1$~mag. We note that all VVV photometric data are on the VISTA magnitude system.


\subsection{Current Status}\label{sec:status}
The VVV Survey started its data-gathering phase in February 2010, with the
first semester focusing on complete multi-band coverage of the whole survey area
(562~square degrees) in its 5 broadband filters (see \S\ref{sec:over}). 
This was completed in 2011. The $K_s$-band observations that comprise the 
VVV variability study started in parallel to the multi-color observations 
in 2010, and are still ongoing. Once completed, the bulge region will
contain between 60 and 100 epochs for each tile, whereas the disk
tiles will have $\approx 60$ epochs each. 

The first two VVV public data releases have already become available from ESO,\footnote{{\tt http://archive.eso.org/cms/eso-archive-news/second-vista-public-survey-data-release.html}} and \citet{rsea12} contains a detailed description of the first VVV data release. 
As of this writing, the various bulge tiles have been observed between 31 and
70 times, whereas the disk tiles have been observed 17 to 36 times
(see also Fig.~\ref{fig:epochs-delays}). Only observations which were
carried out under the observing constraints as specified in the survey
proposal are given the status ``Completed.'' 

With respect to the submitted and scheduled observations, the VVV Survey is $\approx 66\%$ complete. 
Including observations scheduled for years 5 and 6 of the survey, this translates to an overall 
completeness of $\approx 54\%$. Since several VVV tiles have already been observed a few dozen 
times, a first look at stellar variability in the VVV data is already possible~-- and this is 
precisely the main goal of this paper.

\section{Stellar Variability in the VVV Survey}\label{sec:variab}

\subsection{A Brief Overview, and Current Status}
\label{sec:var-over}

The VVV survey has been monitoring the bulge and the southern disk in the $K_s$-band since 2010. It will provide, for the first time, a homogeneous database with long-baseline time-series photometry with up to $100$ epochs for close to $10^9$ point sources. A brief overview of stellar variability in the VVV data was provided in \citet{idea13}. At its current status, after the extensive monitoring of the bulge fields has started, VVV has already provided a considerable number of epochs, suitable for analyses of stellar variability. VVV provides a sparse sampling of the time-domain, usually a single epoch for a few fields on a night (with an occasional second visit), distributed close to randomly over the seasonal visibility period of the area. Figure~\ref{fig:epochs-delays} shows the time sampling of bulge and disk fields. Most of the currently available time-series data (see \S\ref{sec:status}) were taken in the third year, and epochs for some tiles have a moderate clustering in the time sampling. The reason for the similarity in the sampling between different tiles is that observations are carried out in concatenations of nearby fields, in order to provide information on the near-IR atmospheric foreground emission, as is necessary in the data reduction process.

\begin{figure*}
\centering
\includegraphics[width=0.75\textwidth]{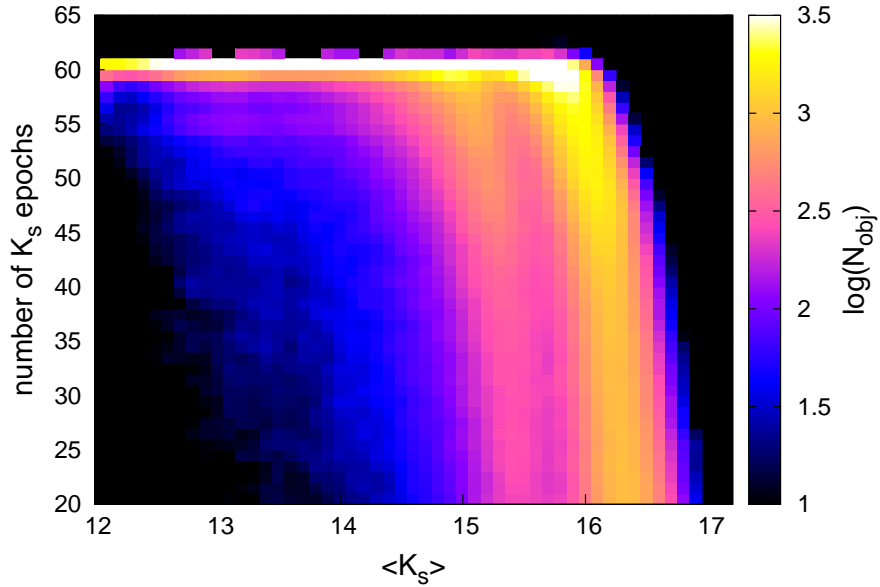}
\vskip0pt
\caption{Distribution of the number of $K_s$-band epochs (after a general threshold rejection procedure) as a function of the average $K_s$ magnitude for objects in VVV bulge field b293, in the data used in \S\ref{sec:var-over}.}
\label{fig:epochs}
\end{figure*}

In the following, we provide a brief overview of the basic properties of the $K_s$-band light curves based on the data of the bulge tile b293 ($\ell=2.3295^{\circ}$, $b=-3.2282^{\circ}$). This tile has 64 epochs with available VDFS photometry, and lies over a moderately crowded stellar field, allowing us to give the current best assessment of the quality of the time-series data. The limiting magnitude in the $K_s$-band varies between $\sim18$ and $\sim16.5$~mag, depending on the Galactic latitude, due to differences in extinction and crowding \citep{rsea12}, and also showing strong nightly variations, mainly due to the highly variable brightness of the near-IR sky, and partly because of the variable seeing or/and cloud coverage during observations. We note that these conditions also cause significant variations in the photometric properties of bright objects, due to the changing saturation level. Figure~\ref{fig:epochs} shows the number of $K_s$ detections for VDFS 1.2 photometry as a function of the $K_s$-band brightness for objects fainter than $12$~mag in field b293. The drop beyond $\sim16$~mag is due to the non-detection of faint objects when observing conditions are sub-optimal. The overwhelming majority of stars brighter than $16$~mag are detected under almost all observing conditions. Most of the bright objects with fewer than 60 detections are crowded and suffer from seeing-dependent merging with nearby sources, while some epochs for the faintest objects are missing mostly due to the high variation in the limiting magnitude caused by the variable near-IR atmospheric foreground. The secondary falling ridge starting at $\sim15$~mag is due to the lower sensitivity of certain VIRCAM chips.

\begin{figure*}
\centering
\includegraphics[width=0.65\textwidth, angle=-90]{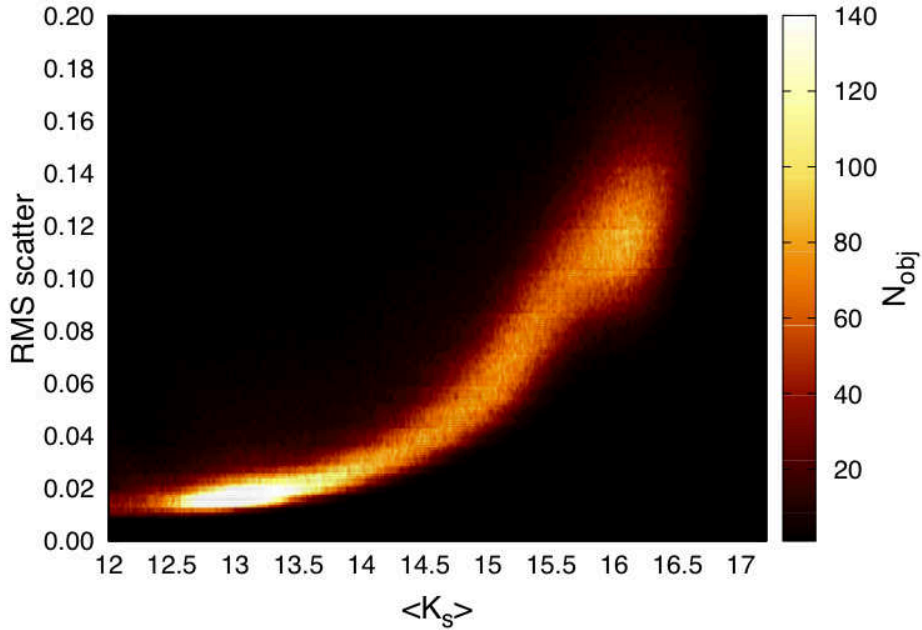}
\vskip0pt
\caption{RMS scatter of the threshold-rejected $K_s$-band light curves of objects in VVV bulge field b293, as a function of the average $K_s$ magnitude.}
\label{fig:sigma-ks}
\end{figure*}

Figure~\ref{fig:sigma-ks} shows the density distribution of VVV variable objects on the root-mean-square (RMS) scatter~--\,average $K_s$ magnitude plane for sources on tile b293 with at least 20 epochs, after applying a general $5\sigma$ threshold-rejection procedure to the light curves. Note that the $K_s$ limiting magnitude of this field is about 0.5~mag fainter than the visible limit of the distribution in Figure~\ref{fig:sigma-ks} \citep{rsea12}, but the majority of the faintest objects was detected only at a few epochs. Figure~\ref{fig:sigma-ks} demonstrates well the high quality of the VDFS photometry~-- note that the photometric ZP is calibrated at each epoch, thus the scatter shown contains not only the formal photometric errors, but also the dispersion from the uncertainties of the nightly ZPs. The low noise at faint magnitudes ($\sim0.1$~mag at $K_s=16$~mag) will allow us to detect RR~Lyrae stars even beyond the bulge. For bright stars ($K_s\sim13$~mag), the photometric precision is around $0.01\,-0.02$~mag, which will allow us to study low-amplitude variables, including planetary transits around nearby K and M dwarfs \citep{rsea11a}, and to investigate the detailed near-IR light curve properties of pulsating and chromospherically active stars alike.

\subsection{Analysis Techniques}\label{sec:techn}

There are various strategies for improving the photometric quality of the VVV light curves with respect to the VDFS catalogs (see \S\ref{sec:var-over}), by means of more sophisticated source extraction and photometric algorithms which may be better suited for dense stellar fields. The VVV Science Team is developing highly automated software pipelines which implement both of the two main approaches that currently deal with crowded-field photometry: PSF-fitting photometry (e.g., DAOPHOT/ALLFRAME, \citealt{pbs87,pbs94}; DoPhot, \citealt{psea93,agea12}) and difference image analysis (DIA; e.g., \citealt{tc96,ala00,db08,jqea10}). One of these pipelines, based on DAOPHOT, is described in \citet{fmea13}. Ultimately, some of these codes will be incorporated into the VDFS, providing a highly robust and versatile data analysis pipeline, equally well suited to handle all types of target fields.


\subsection{Difference Image Analysis}\label{dia}
In addition to using the VDFS photometry catalogs from CASU and DAOPHOT/ DoPhot pipelines (see \S\ref{sec:techn}), VVV is also developing its own DIA pipeline. DIA is particularly effective for obtaining photometry of faint objects in highly crowded fields. We employ a modified version of \citeauthor{ala00}'s optimal image subtraction package ISIS \citep{ala98,ala00}. The standard approach taken by ISIS is to convolve and photometrically match a reference image with good seeing to a target image of poorer seeing in order to obtain a difference image showing only flux changes between the two epochs. A light curve can then be obtained through photometry of a time series of difference images. The advantage of DIA for variability studies is that the difference images contain only varying objects and are therefore unaffected by blended non-varying sources. The convolution kernel to transform the reference image PSF to that of the target image is determined through a least-squares minimization using corresponding sub-regions of the two images.

\begin{figure}
\centering
\includegraphics[width=0.8\textwidth]{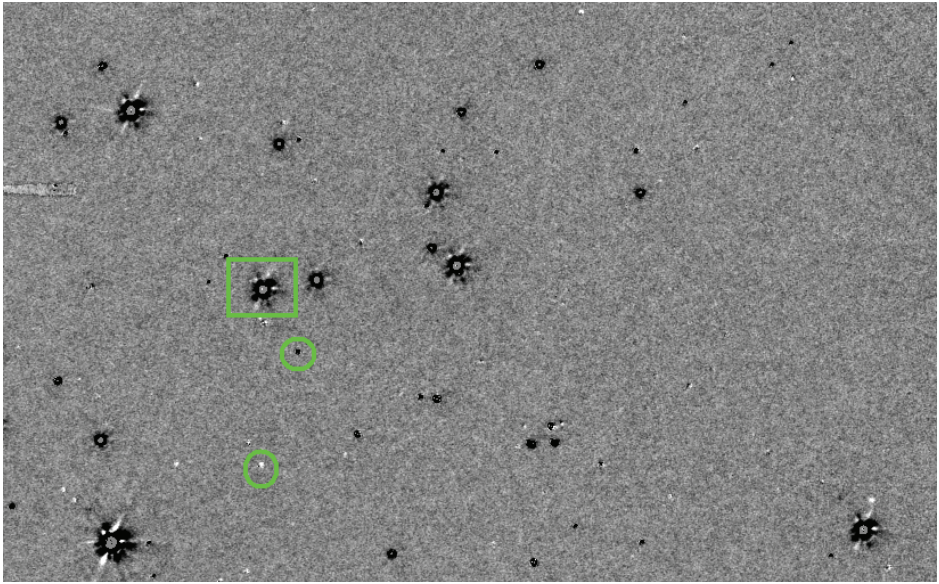}
\caption{A $4\farcm5\times 2\farcm8$ difference image sub-region. Most of the image is gray, reflecting non-varying regions of the image. Varying regions show up as white (black) for flux increases (decreases). Saturated stars typically show up as large blackened structures, sometimes with masked central regions as shown by the example within the rectangle. Genuine variations have a more compact PSF such as the two circled examples.}
\label{fig:dia}
\end{figure}

Whilst the standard approach is to convolve a good seeing reference image to a poorer seeing target image, this may not work well when the PSF is undersampled on the reference image. Ideally, good difference image quality requires typically at least 2.5~pixels/FWHM (i.e., $0\farcs85$ seeing, at the average VIRCAM pixel size), whereas the seeing at the VISTA site in Cerro Paranal is often below this~-- the best-seeing images may have less than 2~pixels/FWHM. For this reason, we choose to use a relatively poor-seeing image with a well-sampled PSF as the reference image, and then convolve a target image with superior seeing to it. For a reference image $R$, a geometrically registered target image $T$ and convolution kernel $K$, we minimize the sum

\begin{equation}
    D^2 = \sum _i [R_i-(T\otimes K)_i + B_i]^2 \,\, , 
\label{newdia}
\end{equation}

\noindent where the sum is over image pixels $i$ and $B$ represents a smooth differential background model. \citeauthor{ala00}'s ISIS package minimizes $D$ with respect to a set of linear coefficients to basis functions which represent both the convolution kernel $K$ and the differential background model $B$. For $K$ the basis functions are a superposition of Gaussian-like functions, and for $B$ the basis functions are 2D polynomials. Figure~\ref{fig:dia} shows an example of a subtracted image that was obtained following this approach. Whilst using a poor-seeing image for the reference may result in a difference image with reduced signal-to-noise ratio, this can be compensated by the fact that all difference images will have the same PSF (unlike \citeauthor{ala00}'s original method) and one can iterate the procedure using the stack of difference images to produce a new, high signal-to-noise reference image.
The VVV DIA pipeline uses a modified version of ISIS which uses equation~(\ref{newdia}) in place of the standard approach. 

Once a difference image sequence has been produced, variable objects are identified from a squared stack of difference images, and PSF photometry is performed for all epochs at the locations of all variable objects (defined above some signal-to-noise threshold). Resulting raw light curves are written to files. We can search these files using specific criteria for e.g. periodic variability, or use some sutiable filter for specific transient objects such as microlenses. At present we are performing searches for regular periodic variables in order to calibrate the performance of the pipeline and, ultimately, catalog target populations such as RR~Lyrae stars. Using DIA we should be able to extend the sensitivity of traditional photometry offered by the VDFS catalogs and detect much fainter variables and transients, as well as variable sources that are blended and thus missing from the CASU catalogs.

\subsubsection{DIA Cross-Matching and Magnitude Calibration}
DIA provides only {\em relative}\/ photometry for variable light curves~-- it does not tie the photometry onto a calibrated magnitude scale. However, absolute calibration for a DIA photometry time-series reduces to calibrating just one epoch as all other epochs are then calibrated through the DIA relative photometry.

We calibrate our DIA photometry using VDFS aperture photometry catalogs for one of the best-seeing epochs in the time series. We refer to this epoch as our {\em photometric reference}\/, which is generally different to the reference image used for image subtraction discussed in \S\ref{dia}. As we have seen, the image subtraction reference image is typically one of poor seeing, in order to maximize sampling of the PSF and ensure good image subtraction. For photometric calibration, however, we choose a good-seeing image, as this provides the best signal-to-noise photometry and minimizes blending effects.

\begin{figure}
\centering
\includegraphics[width=0.8\textwidth]{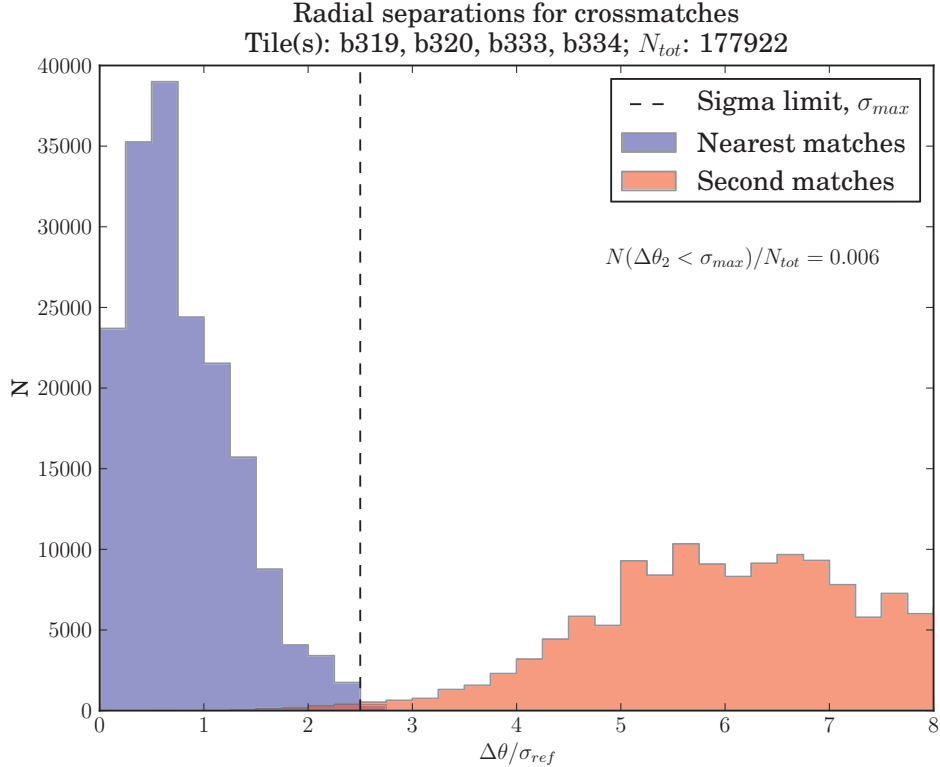}
\caption{Histogram of radial distances for first- and second-nearest VDFS catalog matches to DIA objects for four Galactic center tiles b319, b320, b333, and b334. The total number of objects $N_{\rm tot}$ is given above the plot. The cross-match maximum radius is shown as a dashed line. The first- and second-nearest matches are shown as the blue and red histograms, respectively. The radial distances, $\Delta \theta$, are given in units of $\sigma_{\rm ref} = \theta_{\rm ref}/(2 \sqrt{2 \ln{2}})$, where $\theta_{\rm ref}$ is the DIA reference image seeing. The fraction of potential contaminants (second-nearest matches) within the maximum radius, $N(\Delta \theta_2 < \sigma_{\rm max})/N_{\rm tot}$, is shown beneath the legend.}
\label{fig:xmatchhist}
\end{figure}

In order to tie in the DIA photometry catalogs to the VDFS photometry catalogs we have to cross-match them. Prior to cross-matching the catalogs are first filtered for unsaturated objects which are classified as ``stellar'' by the VDFS pipeline. The RA and Dec coordinates for objects in the DIA source lists are obtained using the World Coordinate System information in the DIA reference image. Each object is then cross-matched to the VDFS object with the smallest angular separation from the DIA position, within a maximum radius of $2.5\,\theta_{\rm ref}/\sqrt{2\,\ln 2}$, where $\theta_{\rm ref}$ is the DIA reference image seeing. The factor of 2.5 was chosen by inspection of a histogram of first- and second-nearest matches for some of the most crowded fields, assuming the second-nearest matches are tracers of potential contaminants. Figure~\ref{fig:xmatchhist} shows an example histogram of radial distances for first- and second-nearest VDFS catalog matches to DIA objects within the four Galactic center tiles.

The VDFS pipeline at CASU performs crowded-field aperture photometry \citep{irwin1985}. For each object, flux counts are obtained for a series of aperture sizes, and used to build a curve-of-growth model. This model is used as an approximation to a PSF profile to calculate the necessary corrections for the conversion of aperture flux to object magnitude. For each VVV DIA object and cross-matched VDFS catalog object, it is necessary to select the largest aperture which contains the majority of the PSF while minimizing the risk of contaminating flux from blended stars. 

The largest VDFS aperture useful for point-source measurements has a
radius of 4~arcsec, which is large enough to guarantee encircling at
least 99\% of the total flux under all expected seeing conditions.
However, within highly crowded fields such an aperture may be strongly
affected by blending, and under typical seeing 99\% of the flux will be
encircled by much smaller apertures. Within the range of 6 VDFS
apertures up to this largest aperture, we select the largest aperture
for which blending effects remain small and we perform this selection
for each object individually.
To this end we compute the forward difference magnitude between neighbor VDFS apertures (which have curve-of-growth aperture corrections applied to them),
$dm_a = |m_{a+1} - m_a|$, and we then determine the aperture $a$, at which $dm_a$ exceeds some limit. This is currently set to $dm_a^{\rm max} = 0.05$~mag, based on an inspection of the aperture magnitude series of a few thousand objects in the Galactic center tiles, but we are seeking a more robust way of determining this limit.

Having established the DIA object magnitude at the photometry reference epoch, $m_{\rm phot}$, we can then transform the difference flux measurements for all epochs onto a calibrated magnitude scale. The magnitude $m_i$ at each epoch $i$ is calibrated from the difference flux, $\Delta F_i$, and the difference flux at the photometry reference epoch, $\Delta F_{\rm phot}$. Taking these together with the (optimal) aperture flux (corrected to the full PSF flux via the curve-of-growth model), $F_{\rm phot}$, the magnitude ZP, $m_0$, and exposure time, $t_{\rm exp}$, from the VDFS cross-matched object, we have:

\begin{equation}
\label{eq:epochmag}
m_i = m_{\rm phot} - \Delta m_{\rm phot} + \Delta m_i  \,\, ,
\end{equation}

\noindent where

\begin{equation}
\label{eq:photmag}
m_{\rm phot} = m_0 - 2.5 \, \log_{10} \left(\frac{{F_{\rm phot}}}{t_{\rm exp}}\right) \,\, ,
\end{equation}

\begin{equation}
\label{eq:photdmag}
\Delta m_{\rm phot} = -2.5 \, \log_{10} \left(
									\frac{F_{\rm phot}}{F_{\rm phot} - \Delta F_{\rm phot}}
								 \right) \,\, ,
\end{equation}

\noindent and 

\begin{equation}
\label{eq:epochdmag}
\Delta m_i = -2.5 \, \log_{10} \left(
		   				\frac{F_{\rm phot} - \Delta F_{\rm phot} + \Delta F_i}{F_{\rm phot} - \Delta F_{phot\rm }}
		   					\right) \,\, . 
\end{equation}

The random error on each individual epoch magnitude is simply the error on its difference magnitude $\sigma_{\Delta m_i}$.
The systematic error in the baseline is obtained from the quadrature sum of the errors in the photometric reference epoch magnitude and difference magnitude:

\begin{equation}
\sigma_{\rm sys}^2
	=	\sigma_{m_{\rm phot}}^2 +
		\sigma_{\Delta m_{\rm phot}}^2. \label{sys}
\end{equation}

\begin{figure}
\centering
\includegraphics[width=0.75\textwidth]{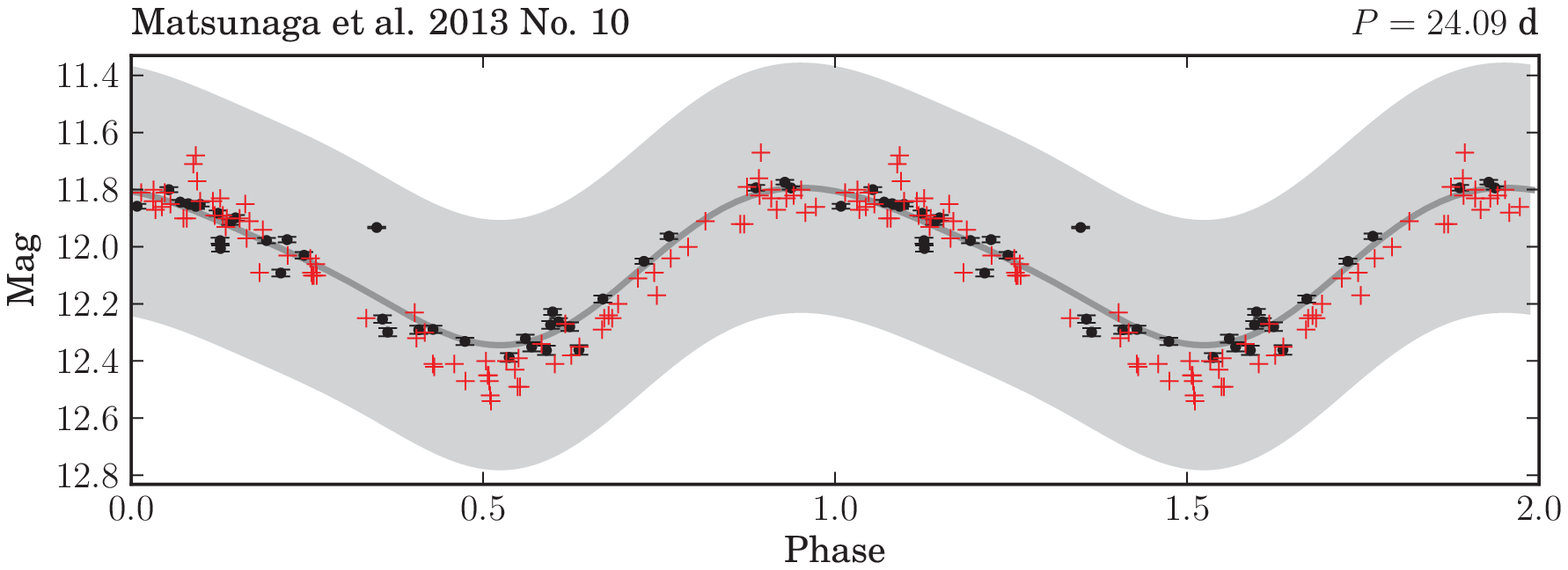}
\includegraphics[width=0.75\textwidth]{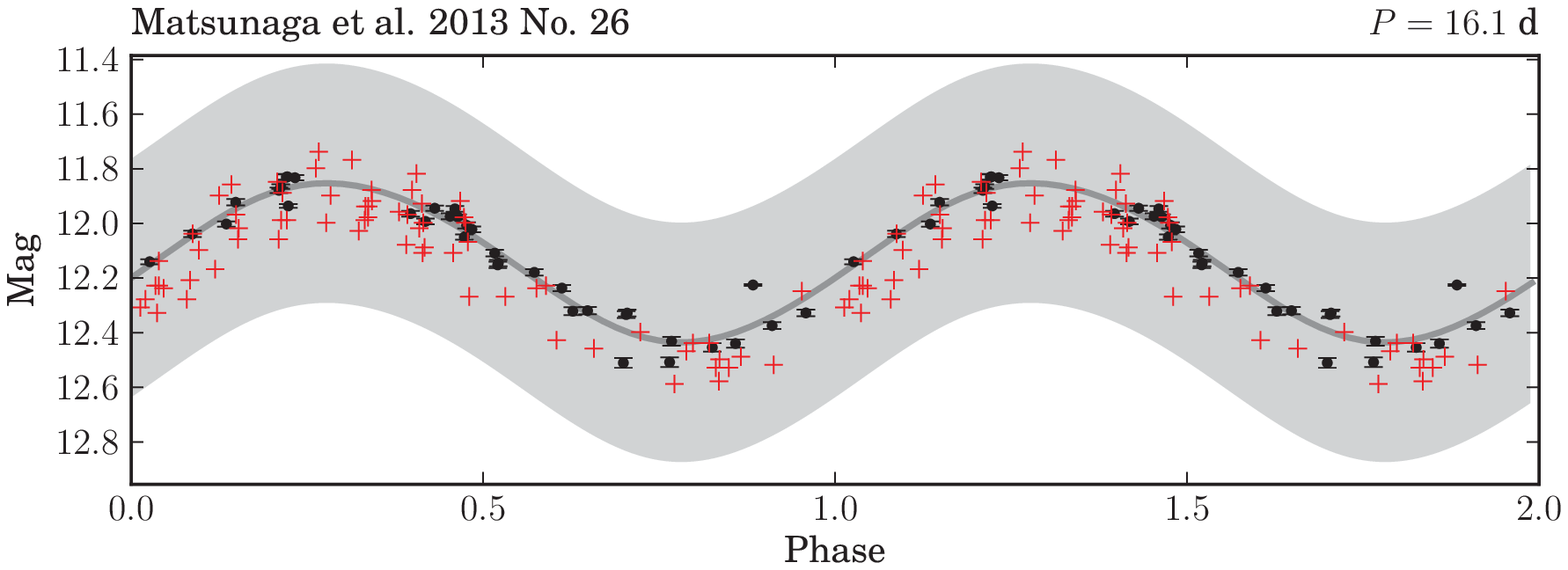}
\caption{Calibrated VVV DIA $K_s$-band light curves for two example Cepheids (Nos. 10, {\em top panel}, and 26, {\em bottom panel}) in \citet{matsunaga2013}. The gray line is a smoothed radial basis function approximation to the VVV DIA data (black points), with the light gray region representing the maximal systematic offset error in the light curve baseline (equation~\ref{sys}). The $K_s$ light curve of \citet{matsunaga2013} (red points) is also shown, offset-corrected to the VVV magnitude calibration for ease of comparison.}
\label{fig:matsunaga}
\end{figure}

The errors $\sigma_{m_{\rm phot}}$, $\sigma_{\Delta m_{\rm phot}}$, and $\sigma_{\Delta m_i}$ are all obtained from equations~(\ref{eq:photmag}-\ref{eq:epochdmag}) through standard propagation of errors.

\begin{figure}
\centering
\includegraphics[width=0.75\textwidth]{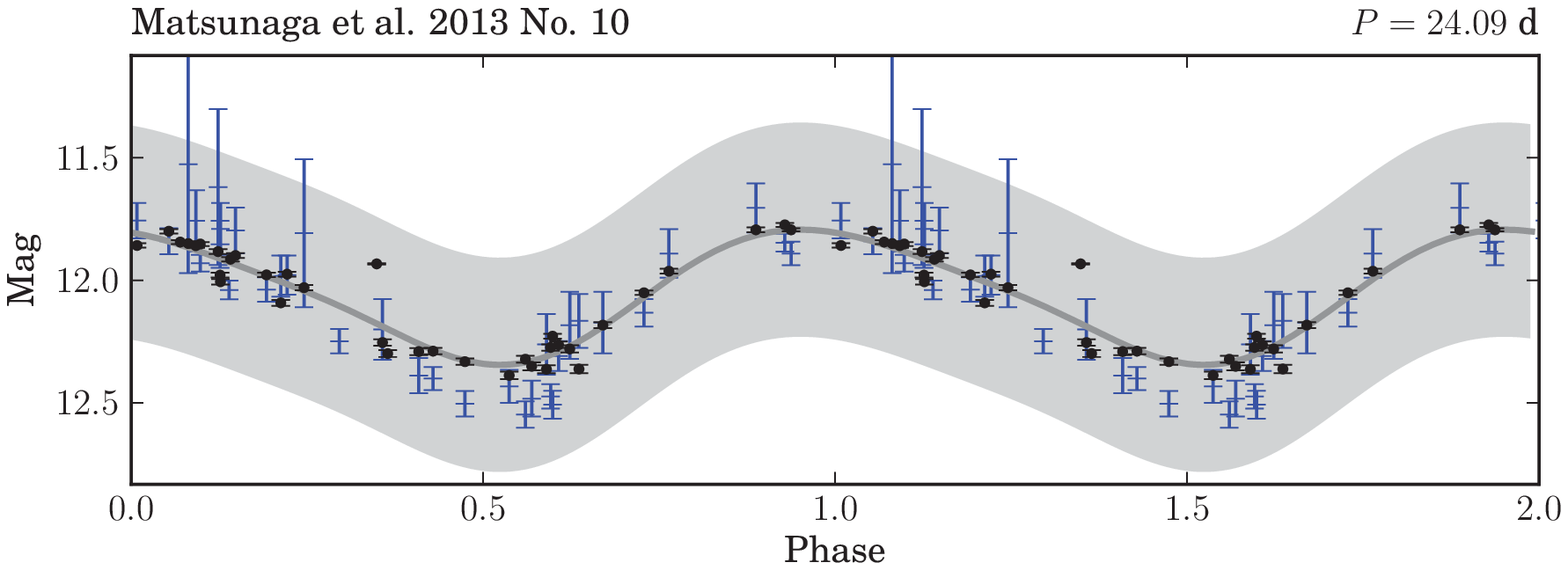}
\includegraphics[width=0.75\textwidth]{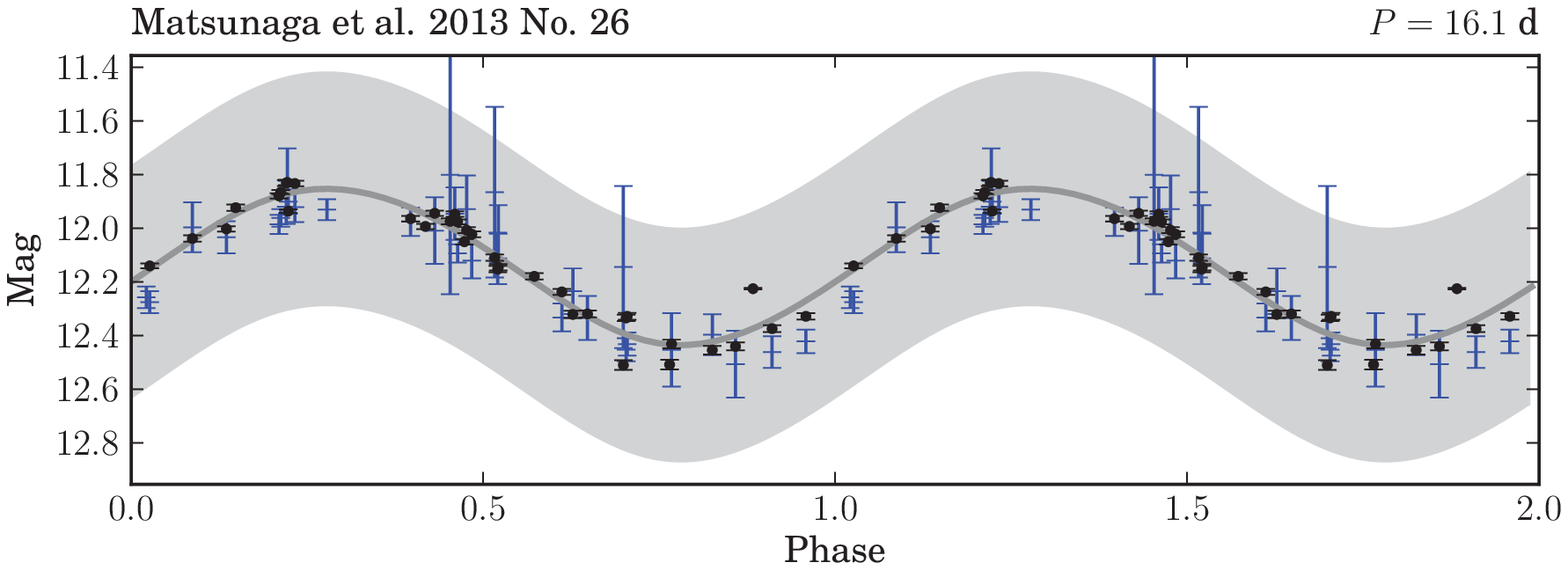}
\caption{Calibrated VVV DIA $K_s$-band light curves for the same Cepheids as in Figure~\ref{fig:matsunaga}, compared with light curves obtained directly from VDFS aperture photometry (aperture number 3; blue points) and offset-corrected. The DIA light curves are identical to those in Figure~\ref{fig:matsunaga}.
}
\label{fig:matsunaga2}
\end{figure}

Figure~\ref{fig:matsunaga} shows calibrated VVV DIA light curves for two Cepheids found by \citet{matsunaga2013}, with their $K_s$-band photometry also shown for comparison. We typically find an offset between the calibrated VVV DIA photometry and the \citeauthor{matsunaga2013} calibrated photometry, though it is within our computed maximal systematic uncertainty (determined from equation~\ref{sys} and shown by the light gray region). In Figure~\ref{fig:matsunaga} we have subtracted off these offsets and we have folded the VVV photometry to the \citeauthor{matsunaga2013} periods for direct comparison. The gray lines represent smooth radial basis function approximations to the VVV DIA data (black points). Figure~\ref{fig:matsunaga2}, in turn, compares VVV DIA photometry with VDFS aperture photometry (aperture number 3) for the same \citeauthor{matsunaga2013} objects as in Figure~\ref{fig:matsunaga}. VDFS aperture photometry also typically shows an offset (subtracted off in Figure~\ref{fig:matsunaga2}) within the systematic uncertainty of the DIA, for reasons which we are still investigating. The random errors from DIA compare favorably with those from aperture photometry.

\subsection{Completeness, Detection of Variable Stars}\label{sec:cross}

The VVV Survey aims at reaching a highly complete census of variable stars that can be used as population tracers, such as radially pulsating stars and eclipsing binary systems. While the old and metal-poor RR~Lyrae stars will be used to trace the 3D structure of the bulge, classical Cepheids and eclipsing binaries will be employed to map the spiral arm structure on the (largely unexplored) far side of the Galaxy, both in the disk region and behind the bulge. In order to get an unbiased picture on Galactic structures traced by these stars, both the rate and precision in the detection of periodic signals have to be sufficiently high. The time-domain coverage at the present stage of the VVV Survey is not yet sufficiently large for conducting variability searches at high completeness, therefore our current investigations are limited to already known samples of distance indicators (\citealt{khea13,idea13b}). However, we are already in the position to give estimates on the future  signal detection performance in VVV based on the time-series data of the 8 VVV bulge fields with more than 60 epochs, which have a partial overlap with the highly complementary optical time-domain survey OGLE-III \citep{is09,so11}.

\begin{figure*}
\centering
\includegraphics[width=0.8\textwidth, angle=-90]{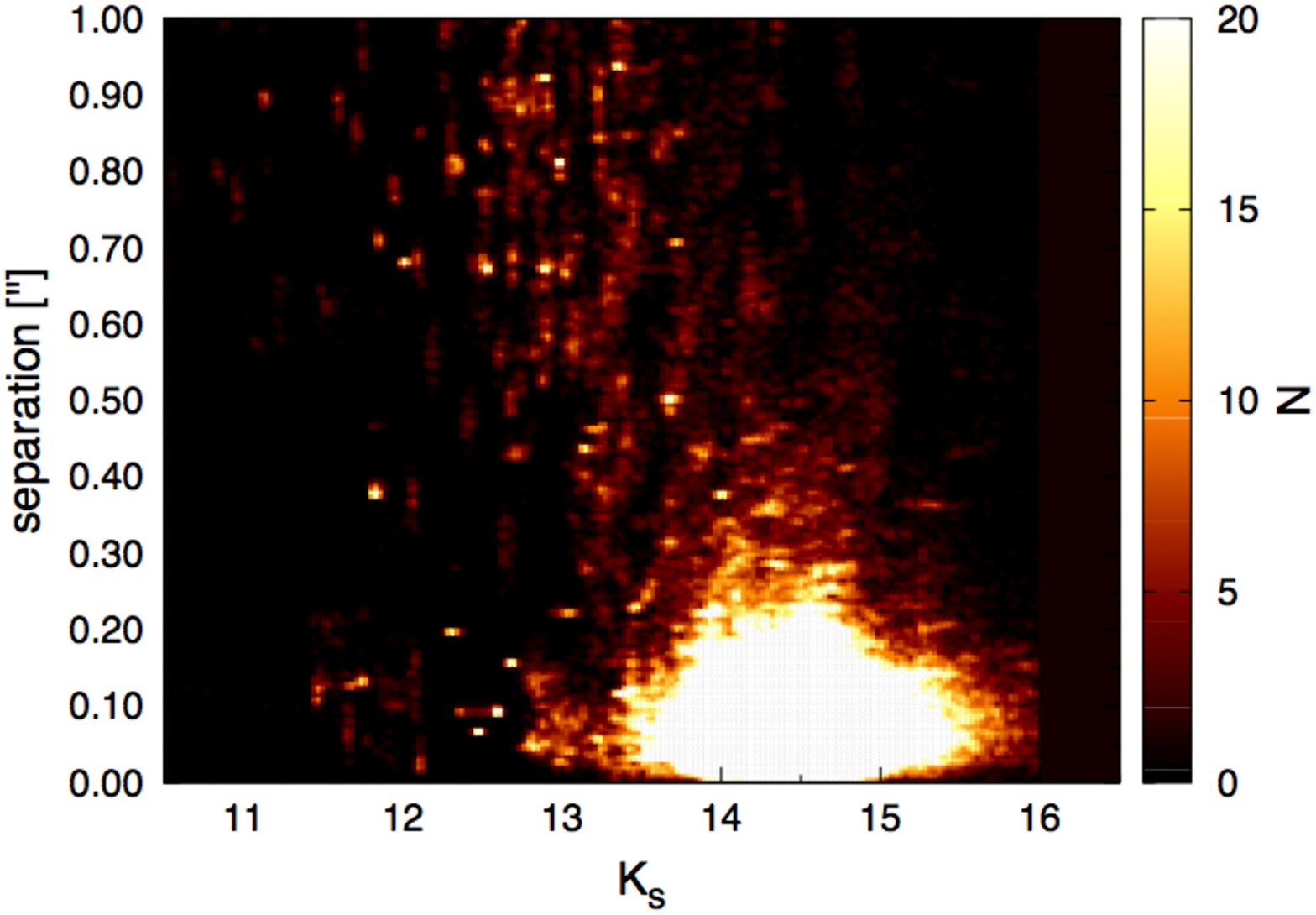}
\vskip6pt
\includegraphics[width=0.8\textwidth]{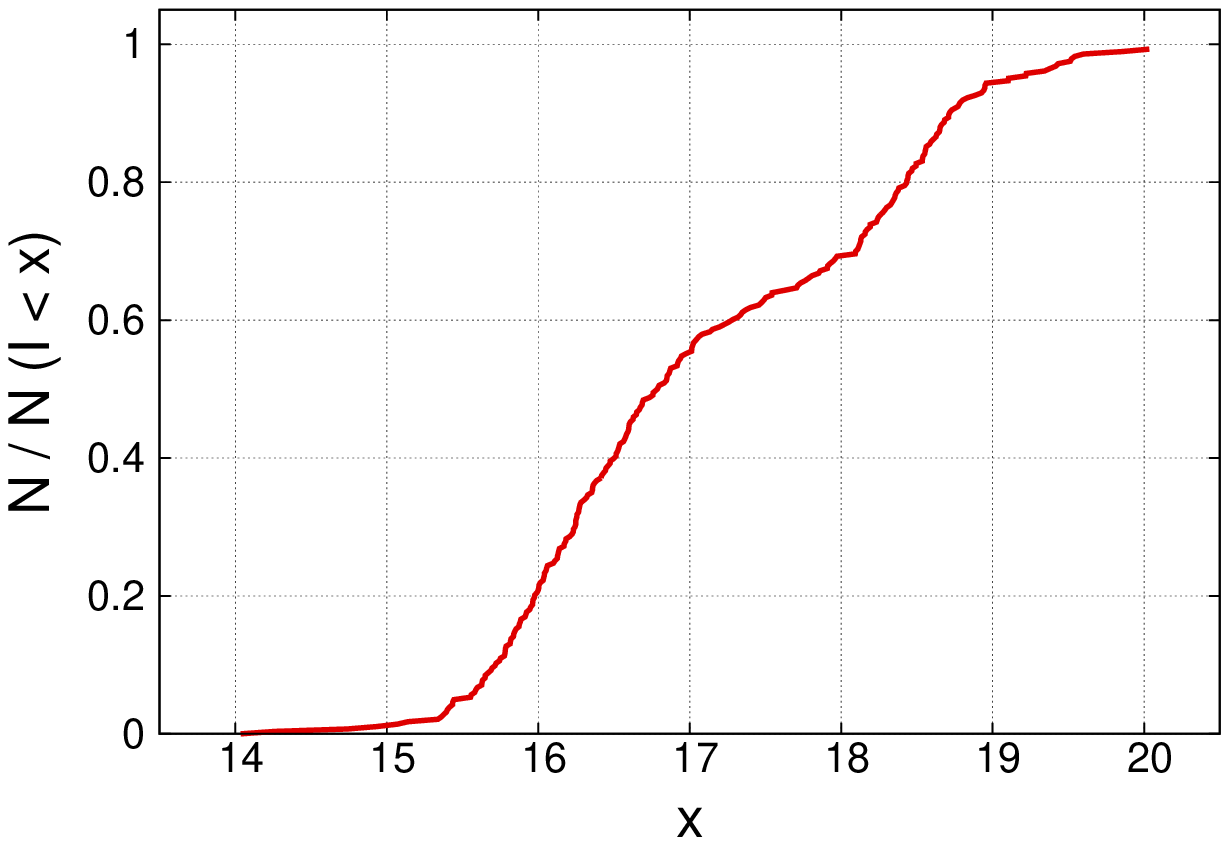}
\vskip0pt
\caption{{\em Top:} number density distribution of the angular separations (in arcseconds) as a function of VVV $K_s$ magnitude of known OGLE-III bulge RRab stars with VVV counterparts in the VDFS source catalogs, located on the 8 VVV fields with the highest number of $K_s$ observations (individual epochs are shown). {\em Bottom:} Cumulative $I$ magnitude distribution of those OGLE-III RRab stars located in the same fields that were not detected in $K_s$ VVV images by the VDFS (i.e., missing from the top panel).}
\label{fig:ksep}
\end{figure*}

In the following, we present a simple and preliminary assessment of signal detection rates based on the data for the ab-type (i.e., fundamental-mode) RR Lyrae stars in this region that are known from OGLE-III. We performed a positional cross-matching procedure between the OGLE-III catalog of RR~Lyrae stars \citep{so11} and the VDFS {\em pawprint} photometric catalogs, including data up to 2012 November (61 epochs). Figure~\ref{fig:ksep} ({\em top panel}) shows the density plot of the angular separations between the best-matching sources as a function of the average $K_s$ magnitude, for all the images. The average cross-matching accuracy is $\sim0\farcs1$, and more than $99\%$ of the data points are concentrated in a small locus with separations less than $0\farcs25$, which indicates that both the precision and accuracy in the astrometry of both surveys is very high (the average VIRCAM pixel size is $0\farcs34$; see \S\ref{sec:over}). The small clusters of points represent complete light curves with poorer cross-matching accuracy, while the points with more diffuse distribution are due to intermittent cross-matching inaccuracies due to, e.g., source merging at poor seeing, elongated sources, spurious signal contamination close to saturated stars, etc. In total, 1558 out of 1832 RRab stars were successfully cross-matched with near-IR counterparts with separations not larger than $0\farcs4$, meaning $85\%$ completeness in on-chip detections with respect to OGLE-III. The main limiting factor for the VDFS photometry is source crowding, since many close objects cannot be separated even using small apertures. PSF photometry and DIA (see \S\ref{dia}) are expected to significantly improve this figure. Figure~\ref{fig:ksep} ({\em bottom panel}) shows the cumulative distribution of average $I$ magnitudes of those RRab stars where near-IR counterparts were not found in VDFS catalogs within the above constraints. The distribution is close to uniform, which means that a similar fraction of objects is missing at bright and faint magnitudes. Thus the limiting magnitude does not affect our completeness compared to OGLE-III. On the contrary, for stars lying close to the Galactic plane, where extinction is severe in optical bands, we expect to have a much higher completeness in the census of variable stars, due to the advantage of the employed near-IR wavebands in penetrating high-extinction regions.

\begin{figure*}
\centering
\includegraphics[width=0.8\textwidth]{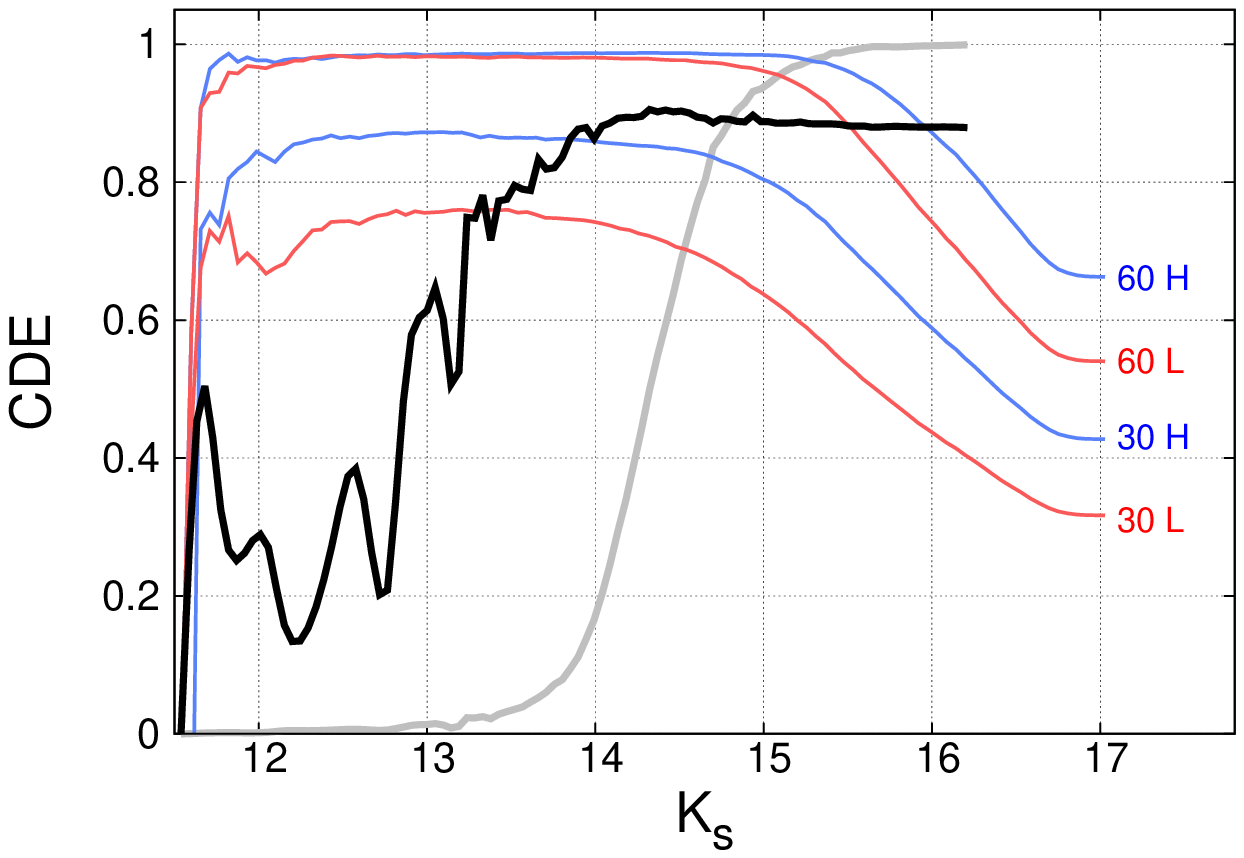}
\includegraphics[width=0.8\textwidth]{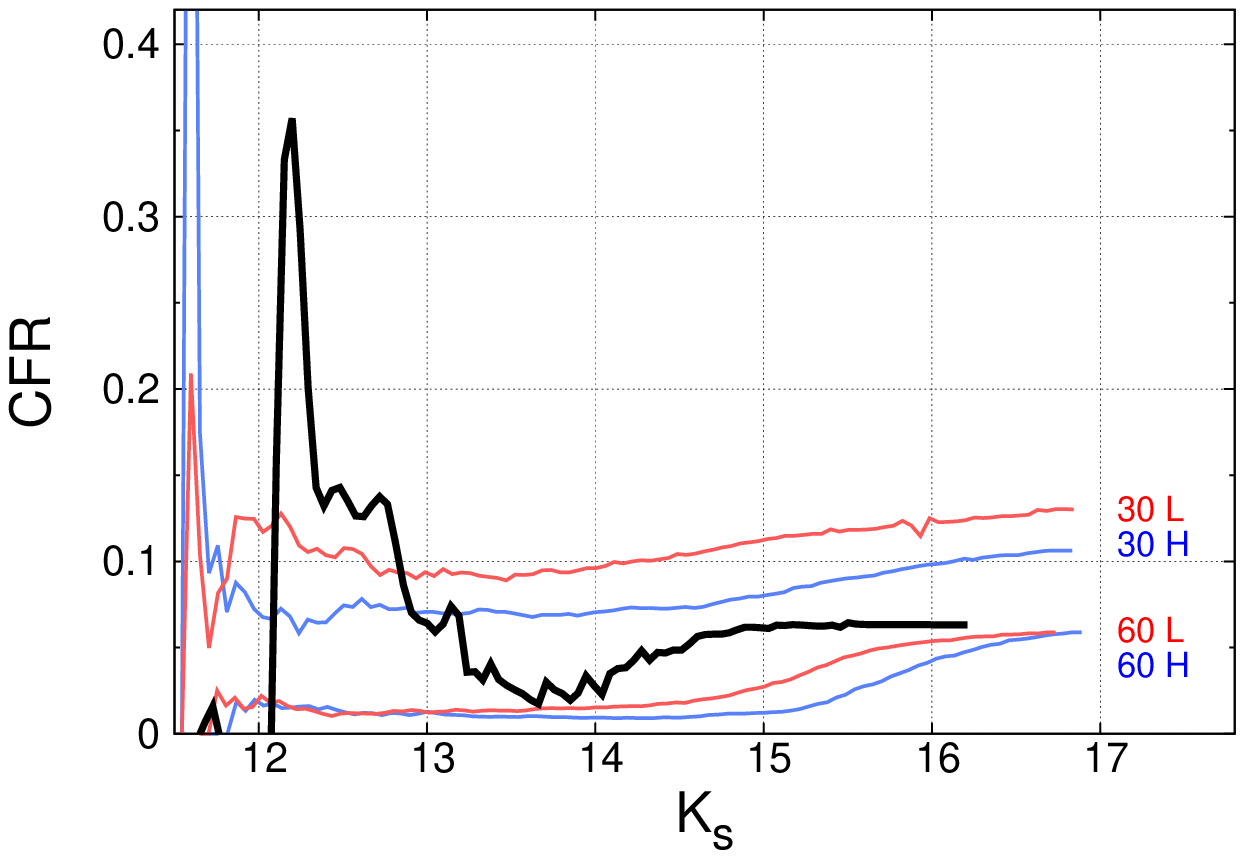}
\vskip0pt
\caption{Cumulative detection efficiency ({\em top}) and cumulative false detection rate ({\em bottom}) of known OGLE-III bulge RRab stars in the VVV Survey with at least $50$ $K_s$ epochs (solid black lines), in comparison with rates measured on simulated low- (L, red) and high- (H, blue) amplitude RRab light curves with 30 and 60 epochs (see text for more details). The gray curve on the top panel shows the cumulative magnitude distribution of the OGLE-III data used in this figure.}
\label{fig:cdp}
\end{figure*}

To evaluate the signal detection rate, we compute the Lomb-Scargle periodograms \citep{nl76,js82} of the VVV $K_s$ light curves of successfully cross-matched RRab stars that have at least 50 epochs (1076 objects). Then, we characterize the completeness in signal detection by the cumulative detection efficiency (CDE):

\begin{equation}
\label{cde}
{\rm CDE} = \frac{N_{\rm det}(m<K_s)}{N_{\rm all}(m<K_s)} \,\, ,
\end{equation}

\noindent where $N_{\rm det}$ is the number of successfully detected RRab stars, and $N_{\rm all}$ is the total number of RRab stars, respectively~-- both including all stars with magnitude $m$ up to a certain magnitude $K_s$. We consider a detection successful if the signal is within the $1\%$ significance level evaluated on white noise, and its frequency, refined by non-linear least-squares fitting, does not differ from the one reported by OGLE-III by more than the nominal frequency resolution (i.e., the inverse of the baseline, or $\sim0.001\,{\rm d}^{-1}$). The result is shown in Figure~\ref{fig:cdp}, compared with detection efficiencies computed from synthetic data. For the latter, a typical $K_s$-band RRab light curve (similar to SW~And; see \citealt{rjea96}, and also Fig.~\ref{fig:felipe} below) with low and high amplitude ($P = 0.87642~{\rm d}$, $A=0.2$~mag; and $P = 0. 44226~{\rm d}$, $A=0.3$~mag, respectively) was sampled randomly within the visibility periods of the bulge in a VVV-like scenario, random noise was added based on the scatter diagram shown in Figure~\ref{fig:sigma-ks}, and the CDE was computed from a few thousand realizations. We conclude that a signal detection rate of $\sim90\%$ can already be reached on VVV data for RRab stars in the bulge, in light curves sampled with $\sim 60$ epochs. This number will be pushed significantly higher by raising the number of epochs to $80-100$ (i.e., increasing the spectral $S/N$ and resolution, as well as decreasing the effect of aliasing), and improving the photometric precision by either profile-fitting photometry or DIA.\footnote{We note that the fluctuations in the CDE at low magnitudes are due to the relatively low number of bright stars in the OGLE-III sample.}

Another important aspect of signal detection efficiency is the relative number of significant signals with incorrect periods, e.g., due to aliasing. Ideally, the number of such cases should not exceed a few percent because, for instance, wrong periods imply wrong distances through the period-luminosity (PL) relations, and if these errors are too frequent, they can blur the 3D structures that we wish to trace. The relative frequency of signals with incorrectly recovered periods can be quantified by the cumulative false detection rate (CFR), defined analogously to CDE (eq.~\ref{cde}):

\begin{equation}
\label{cfr}
{\rm CFR} = \frac{N_{\rm false}(m<K_s)}{N_{\rm sign}(m<K_s)} \,\, ,
\end{equation}

\noindent where $N_{\rm false}$ is the cumulative number of detections where the frequency is incorrect, and $N_{\rm sign}$ is the total number of light curves with significant signals, both again including sources up to a certain magnitude. The top panel of Figure~\ref{fig:cdp} shows the CFR measured in our RRab sample, once again in comparison with the values computed from synthetic data (see above). The CFR is about $5\%$ for bulge RRab stars, which is already sufficient to rather sharply trace the 3D structure of the underlying stellar population, and we expect to reach $1-2\%$ by the end of the survey. We note that the peak close to 12~mag is produced by very few stars, and is caused by heavy aliasing introduced by the intermittent saturation of these bright objects.

\subsection{Automated Classification}\label{sec:class}
Current estimates based on the analysis of the available VISTA datasets as delivered by the CASU VDFS pipeline have suggested that the final number of variable stars observed in the VVV Survey may be in the range between 10$^6$ and 10$^7$ stars. These large numbers of objects require new approaches for the data analysis and selection, including artificial intelligence algorithms. Machine learning techniques applied to variable star classification have become particularly popular in recent years. For instance, in \citet{jdea07} the authors explored several classification techniques, quantitatively comparing performance (e.g., in terms of computational time) and final results (e.g., in terms of accuracy) of different classifiers with their corresponding learning algorithms. More recently, a few other studies have focused on specific methodologies, with the implicit goal of finding the best compromise between robustness and speed. As an example, \citet{pdea11} and \citet{jrea11} have independently presented tree-based methods for the automated classification of Hipparcos and OGLE variable stars; \citet{kpea12} have employed machine learning techniques to detect quasi-stellar objects in the MACHO and EROS-2 databases; while \citet{jbea11} opted for algorithms based on multivariate Bayesian statistics in order to explore the variable star content of the TrES Lyr1 sky field.

Whichever specific algorithms preferred, the general idea behind these supervised machine learning methodologies is to create a function, the \textit{classifier}, able to infer the most probable \textit{label} of an object (in our case, the variability class to which an unclassified variable star belongs to) on the basis of what is learned by the analysis of inputs (light curve features) from a \textit{training set} (a collection of high-quality light curves of previously classified variable stars). In the most general framework, the sequence of steps to be performed can be thought as: 1)~build a training set (template database); 2)~determine the input feature representation (e.g., periods and harmonics, as derived from Fourier analysis) of the learned function;  3)~determine the nature of the classifier with its corresponding learning algorithm (e.g., artificial neural network, support vector machines (SVMs), tree-based methods, etc.); 4)~run the classifier on the gathered training set, using the information on well-known variables stored in the training set for searching and labeling unknown variable stars in the test set, i.e., in the data archive that one is dealing with; 5)~finally, evaluate the accuracy of the learned function, i.e., evaluate the fraction of correctly classified variable stars \citep[e.g., through a so-called confusion matrix;][]{jdea07}. In what follows, we describe in some detail the first three points of this proposed working strategy.

\subsubsection{VVV Templates Database}\label{sec:templates}
Until now, stellar variability in the near-IR has been a relatively ill-explored research field: in particular, the number of high-quality light curves was very limited and, even worse, many variability classes have not yet been observed in a sufficiently extensive way in the near-IR, so that good light curves are entirely lacking for some such classes. Since VVV is the first ever large survey dedicated to stellar variability in the near-IR, the first problem we had to face has thus been the construction of a proper training set, i.e., a database of high-quality (``template'') near-IR light curves taken to be representative of the different variability classes under study. Our effort in building such a template database has included four main routes, which are described in the following.\footnote{See also {\tt http://www.vvvtemplates.org/}.} 

\begin{figure*}
\begin{center}
\includegraphics[width=0.8\textwidth]{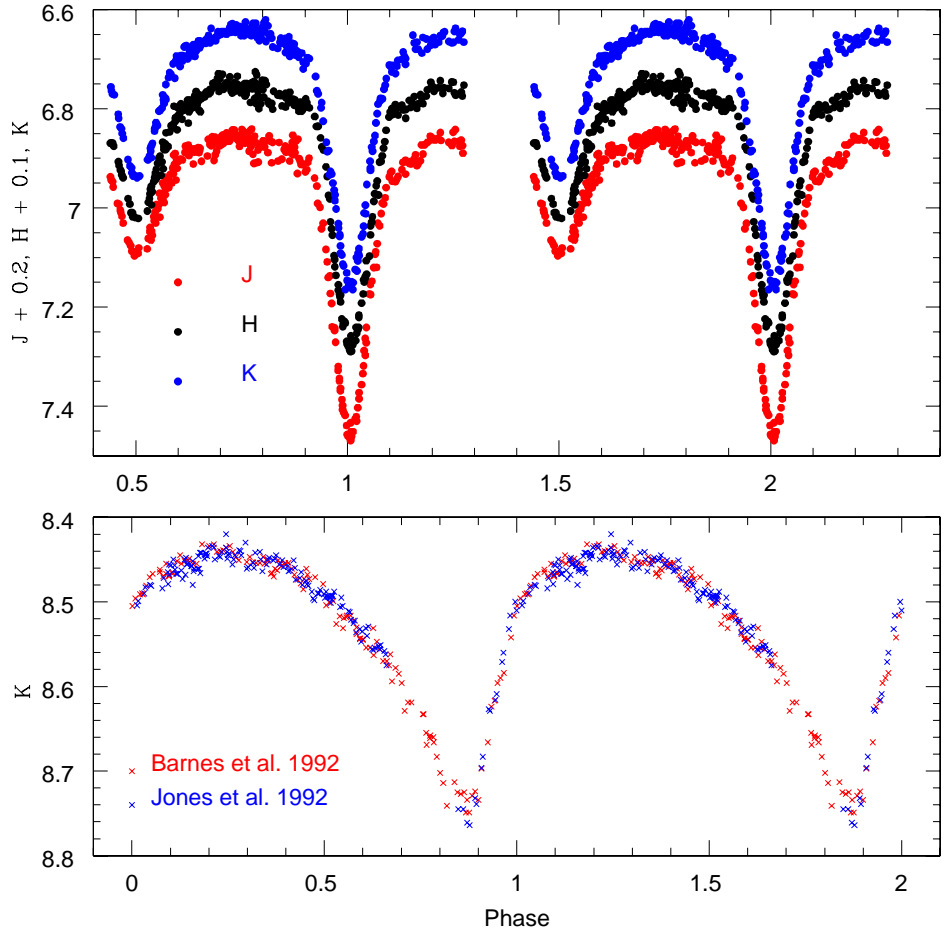}
\end{center}
\caption{An example of data digitalization from the literature. {\em Top panel}: AI Dra ($P = 1.1988146$~d), an eclipsing binary of the
  Algol type, in the $J$ (red), $H$ (black), and $K$ (blue) bands, from \citet{Lazaro04}.
  {\em Bottom panel}: SW And ($P = 0.442$~d), an RRab-type variable. Red crosses are data points from \citet{jones92}, whereas blue ones are from \citet{barnes92}.
\label{fig:felipe}}
\end{figure*}

\smallskip 
\smallskip 
\noindent \textbf{\normalsize\textit{a)~Near-IR Light Curves from the Literature.}} Firstly, we have extensively explored the available literature, searching for papers that presented high-quality near-IR time-series data. When these data were not published in machine-readable form and we could not obtain the data directly from the authors, we digitized the data tables and/or plots, and used optical character recognition software to convert those into ASCII files. Among the types of variables for which data could be retrieved in this way are RR Lyrae, Classical Cepheids, Miras, eclipsing binaries, and Wolf-Rayet stars, among others. Examples of high-quality data obtained in this way are shown in Figure~\ref{fig:felipe}.

\begin{figure*}[ht]
\begin{center}
\includegraphics[width=0.82\textwidth]{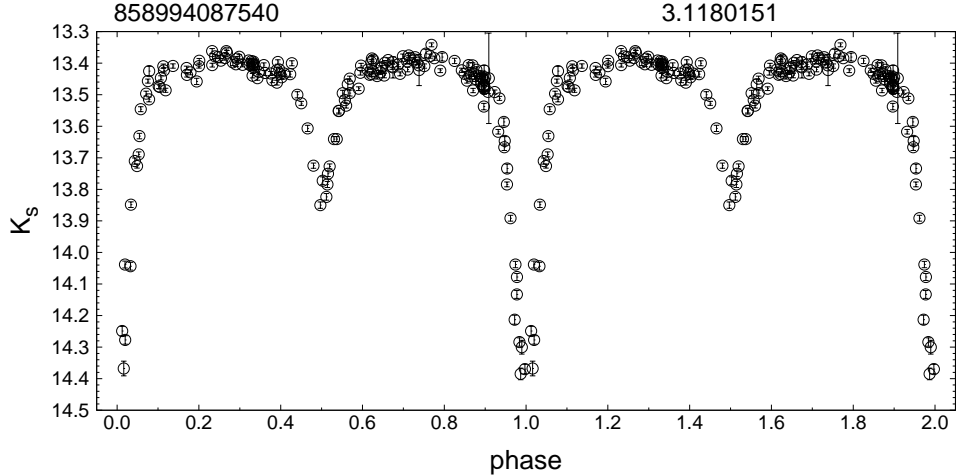}
\end{center}
\caption{$K_s$-band light curve of a 3.1-day eclipsing variable, based on WFCAMCAL data (Ferreira Lopes et al. 2013, in preparation). \label{fig:wfcam}}
\end{figure*}

\smallskip 
\smallskip 
\noindent \textbf{\normalsize\textit{b)~Near-IR Light Curves from Public Archives.}} We have also searched publicly available archival databases for high-quality near-IR data, finding the 3.8m United Kingdom Infrared Telescope (UKIRT) Wide-Field Camera (WFCAM; \citealt{mcea07}) Calibration Archive (WFCAMCAL) especially appropriate for our purposes. WFCAMCAL's current data release (DR8) contains data from 364,905 individual pointings on both the Northern and Southern Hemispheres, spread over nearly half of the sky. The majority of the fields are observed repeatedly, with a rather irregular sampling that has however provided for many sources a reasonably extended time coverage. The selection of variable sources from the WFCAMCAL catalog and the corresponding light curves will be presented in Ferreira Lopes et al. (2013, in preparation); an example of a light curve derived in this way is provided in Figure~\ref{fig:wfcam}.

\begin{figure*}[h]
\begin{center}
\includegraphics[width=0.85\textwidth]{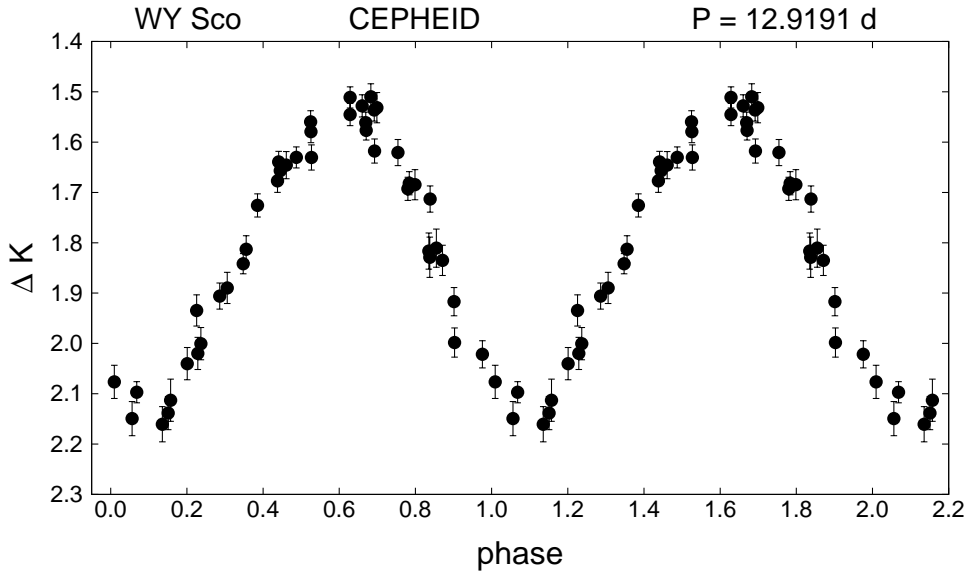}
\end{center}
\caption{Differential $K$-band light curve of the classical Cepheid WY Sco, obtained with the REM 0.6m telescope (Angeloni et al. 2013, in preparation).}\label{fig:wysco}
\end{figure*}

\smallskip 
\smallskip 
\noindent \textbf{\normalsize\textit{c)~VVV Templates Observing Project.}} In addition to data from the literature and from public archives, we have also carried out time-series near-IR observations of our own. In this way, we have monitored hundreds of (optically well-studied) variable stars in the $J$, $H$, and $K_s$ bands, using several different facilities located at different observatories across the globe (see Table~3 in \citealt{mcea11} for a listing of all the telescopes and instruments used). An example of a light curve obtained in this way is shown in Figure~\ref{fig:wysco}; additional examples can be found in \citet{mcea11}. A full description of the project will be provided in Angeloni et al. (2013, in preparation).

\begin{figure*}
\begin{center}
\includegraphics[width=0.75\textwidth]{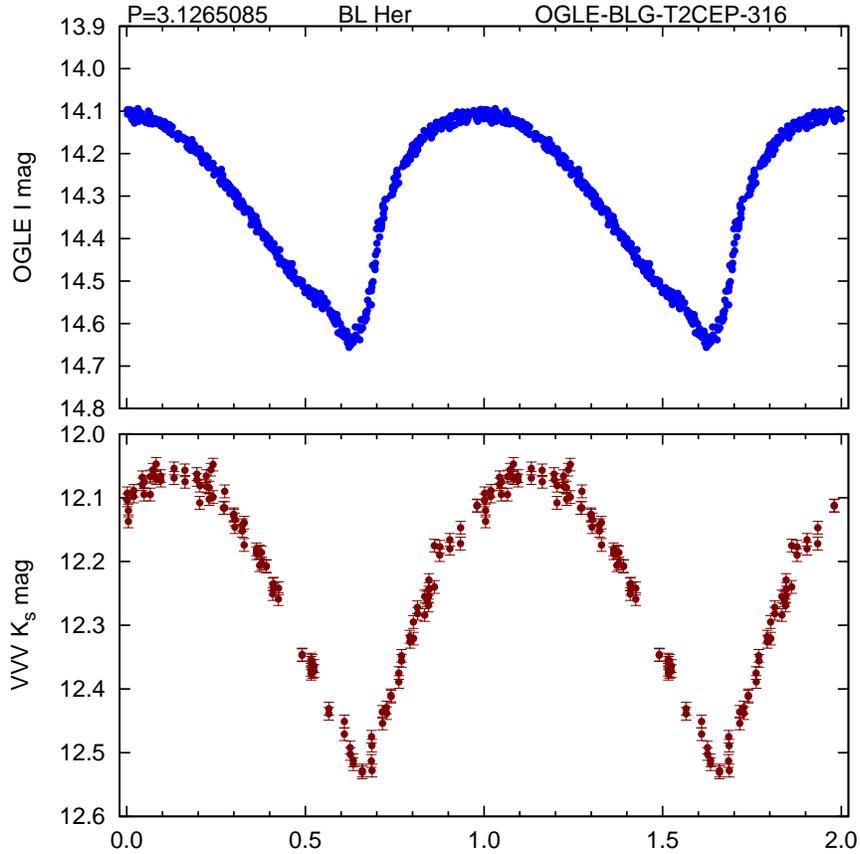}
\end{center}
\caption{Template light curve of a BL Her-type variable in the Galactic bulge. {\em Top panel}: OGLE $I$-band data. {\em Bottom panel}: VVV $K_s$-band data.}\label{fig:vvv_blher}
\end{figure*}

\smallskip 
\smallskip 
\noindent \textbf{\normalsize\textit{d)~VVV Light Curves.}} Last but not least, VVV itself has started delivering light curves of variable stars that had already been classified by previous optical surveys, most notably MACHO \citep{caea95} and OGLE \citep{auea92}. As an example, Figure~\ref{fig:vvv_blher} shows a BL Her (type II Cepheid) light curve, obtained using OGLE-III $I$-band data \citep{isea11,isea13} and VVV $K_s$ data. As more and more light curves of this quality become available, our template database will be augmented accordingly, thus enabling increasingly more accurate classification of previously unclassified VVV variable sources.

\subsubsection{Feature Extraction and Classification of Variable Stars}\label{sec:features}
The problem of automated classification of variable stars using temporal series has been around for some time. For example, \citet{jdea07} have successfully implemented algorithms for automated classification which are based on extensive light curve databases, containing thousands of entries, from Hipparcos, OGLE, and other projects~-- all providing light curves in the visible bandpasses only. In the case of near-IR projects like VVV, special care must be taken both in the feature extraction process and in the implementation of classification algorithms, because the number of training examples (i.e., high-quality light curve templates) that is available at the outset is not as large.
 
How does one optimally identify and extract the most informative features that best describe a temporal series? The usual model for a light curve of a given variable star, where $m(t)$ represents the magnitude as a function of time $t$ (usually Julian Days), is

\begin{equation}
\label{fullfit}
m(t)=a+b\,t+\mathrm{HS}(t)+\varepsilon(t) \,\, ,
\end{equation}

\noindent where

\begin{eqnarray*}
\mathrm{HS}(t)=\sum_{i=1}^{n_f} \sum_{j=1}^{n_i}a_{ij}\sin(2\pi jf_it)+b_{ij}\cos(2\pi jf_it) \,\, . 
\end{eqnarray*}

\noindent Here $a$ and $b$ represent the intercept and slope respectively of the  linear trend in the time series, $\mathrm{HS}(t)$ is the harmonic sum 
(which tries to describe the star's actual variability), $n_i$ is the number of harmonics 
for a given frequency $f_i$, $n_f$ is the number of frequencies obtained from the light curve, and 
$\varepsilon(t)$ is the (possibly correlated) noise. From here,  obtaining  features of the 
light curve is straightforward: the frequencies could be estimated via the ``classical'' 
Lomb-Scargle periodogram \citep{nl76,js82}, first detecting and subtracting the linear trend, 
then detecting and subtracting the frequencies one at a time (i.e., performing prewhitening), and finally 
performing a full least-squares fit with the model in equation~(\ref{fullfit}). From here, one can decide to 
extract some useful features for classification. Following the work of \citet{jdea07}, these features are (i)~the parameters of the linear trend; (ii)~the amplitudes, which are given by 

\begin{equation}
A_{ij}=\sqrt{a_{ij}^2+b_{ij}^2} \,\, ;
\end{equation}

\noindent (iii)~the (time-invariant) phases, given by 

\begin{equation}
\phi_{ij}=\arctan{\left[\sin({\rm PH}_{ij}),\cos({\rm PH}_{ij})\right]} \,\, ,
\end{equation}

\noindent where 

\begin{equation}
{\rm PH}_{ij}=\arctan(b_{ij},a_{ij})-(jf_i/f_1)\arctan(b_{11},a_{11})
\end{equation}

\noindent (with $\phi_{00}=0$); and (iv)~the ratio $\sigma^2_{f_1}/\sigma^2$ of two variances. $\sigma^2$ corresponds to the variance of the raw photometric measurements, while $\sigma^2_{f_1}$ is the variance from the residual \textit{after} the first frequency fit (with the corresponding $n_1$ harmonics of this first frequency). 

The above prescribed procedure, though somewhat ``standard,'' has nonetheless some issues that must be carefully addressed. 

The first issue with which we must be careful is whether classical periodogram searches, such as the Lomb-Scargle method, constitute the optimum solution for the frequency/period searches. For example, \citet{zk09} claimed that the classical periodogram method can be thought of as trying to fit sine curves (with the corresponding phase) of different frequencies, i.e., a fit of the form

\begin{equation}
S(t)=a_k\sin(2\pi f_kt)+b_k\cos(2\pi f_kt) \,\, , 
\end{equation}

\noindent where the periodogram is just a function of the amplitude of this sinusoid for each frequency $f_k$. 
According to these authors, this procedure does not take into account the fact that the sinusoid may be 
``floating'' around a different mean value than that of the linear trend in equation~(\ref{fullfit}). They 
claim, in addition, that this procedure does not weigh the data points according to their respective variances, as they 
should in order to have unbiased estimators for the amplitudes in the presence of uncorrelated noise. To solve 
these issues, \citeauthor{zk09} add an extra term to the function $S(t)$ above to account for this ``floating mean,'' 
and write the least-squares solution assuming a diagonal covariance structure for the data, where each 
entry of this diagonal covariance matrix corresponds to the variance on each data point. 
\citet{rvea10}, on the other hand, disagree with this procedure. They actually claim that performing 
least-square fits to sinusoids is different from calculating the periodogram, which in essence is the 
squared amplitude of the discrete Fourier transform of the data. Moreover, they claim that their 
formalism is capable of dealing with all types of colored noise. 

The second issue that one has to take care of is aliasing, i.e., spurious peaks due to sampling. 
For example, \citet{Dawson10} have shown that the problem of aliasing can trick not only period-finding algorithms, but 
also the human eye, misidentifying a ``real'' peak in the periodogram by a fake peak produced by 
sampling (ir)regularities. This is a very serious and critical problem for classification purposes: 
if we fail to identify the true period of a periodic variable star, then we lose perhaps its most important 
feature. Basically, \citeauthor{Dawson10} showed that it is not always true that the highest peak in the 
periodogram is actually a (physical) period of the object of study.

The third issue is how to select $n_f$, the number of frequencies in the harmonic fit~-- and, given that number, 
how to select $n_i$, the number of harmonics in each frequency fit. Even further, one may ask whether or not the number 
of harmonics should be the same for all the frequencies. Physically, there is no a priori reason to 
assume that the number of harmonics should be the same, nor that all variability types are characterized by the same number of frequencies. 

As can be seen, the issue of feature identification and extraction remains subject to considerable debate.  
We are currently conducting extensive tests using VVV data and our templates database
(\S\ref{sec:templates}), in order to obtain an 
optimum strategy for the purpose of classifying VVV light curves. 

As a first approach to solve some of the issues mentioned above, we
propose to initially over-fit the model in equation~(\ref{fullfit}), using
more features than possibly necessary to fit the light curves. In more
detail, we force our model to fit for each light curve up to five
frequencies and up to five harmonics for each frequency to obtain the
features with which we represent the light curves. Therefore, each
light curve can now be represented as a vector $\mathbf{x}$ containing
the frequencies, amplitudes, phases, slope and intercept from the
linear trend, and ratio of variances. In a second step, we fit a
binary classifier,
$Pr(Y^c=1)=g(\beta_0^c+\beta_1^cx_1+\ldots+\beta_m^cx_m)$, for each
class $c$ which can learn how to discriminate between classes and
perform feature selection {\em simultaneously}. This classifier is
called LASSO. The LASSO classifier estimates the parameters
$\beta_0^c,\ldots,\beta_m^c$ by minimizing a cost function subject to
a constraint on the size of the $L_1$ norm of the parameter vector
given by $\sum |\beta_i|<\lambda$. This constraint on the
size of the parameter vector shrinks the parameter estimates towards
zero, and the use of the $L_1$ norm forces some of the parameters to be
equal to zero. The features associated with parameters that
are estimated as being equal to zero are irrelevant for the
classifier.  Therefore, this procedure allows us to have a different
set of features for each classifier, and is convenient in the present
setting in which we need to classify variable stars which can be
mono-periodic, multi-periodic, or non-periodic. For example,
mono-periodic light curves should need fewer frequencies in the
harmonic fit than multi-periodic stars, which implies that fewer features
are needed for the classifier~-- and this is borne out naturally by LASSO.

\subsubsection{Training Templates and Classifiers}\label{sec:machine}
In order to improve our classifiers as new light curve data become
available, we plan to augment our templates training set (\S\ref{sec:templates}) by
selecting the most informative light curves through an active learning
methodology. Active learning procedures are very helpful to efficiently select the most informative instances to be included in training sets \citep{Cohn1992,Roy:McCallum:2001,Tong:Koller:2001}. In the astronomical context, \citet{Richards:2012ApJ} recently used active learning and showed its benefits for handling astronomical data. These authors show that active learning techniques can reduce the bias of the training process and increase the classification accuracy. Most active learning models are composed of two phases. The first is the exploration phase, where the model explores the most informative instances to select. The second is the exploitation phase, where the model includes the feedback after the new selected instance and updates itself in order to repeat the exploitation phase \citep{Cohn1992,Roy:McCallum:2001,Tong:Koller:2001,Pelleg:Moore:2004,Cebron2008}. Usually the exploration phase requires a lot of computations, in most cases passing through the data many times, making it very difficult to be directly applied to large astronomical catalogs. The exploitation phase may be less costly, depending on the model used. As in \citet{Richards:2012ApJ}, most active learning models assume the existence of an oracle, an entity that can correctly classify any query instance. Unfortunately, such an oracle may not be available, or/and may be too expensive to implement. Trying to deal with the absence of an oracle, semi-supervised active learning techniques attempt to use only the available labels \citep{Zhu03combiningactive,Wu:2005,Ambati:2010}. These models use expectation maximization to estimate the best prediction for the missing labels using the current ones, but unfortunately most of these models cannot handle huge datasets because of the computational cost of the algorithm. Recent work introduced active learning for large datasets \citep{Jain2010}, based on a similar approach as in \citet{Tong:Koller:2001} but using hashing techniques to speed up the process. This provides a very interesting approach to the \lq\lq large datasets" problem, but still assumes the existence of an oracle.

In this context, we are currently  developing an active learning framework  for large datasets, modelling the partial absence of the oracle. After developing this semi-supervised active learning framework, we will iterate between  automated classification of the available VVV light curves and increasing the size of the training set by incorporating new template-quality light curves (i.e., light curves with extremely high classification probability, as judged by the code and/or the oracle, in this case a VVV astronomer) to the latter. This will allow us to refine the classification procedure, and then re-run the whole process with increased confidence as additional data are incorporated into the main VVV light curve database. In this way, our final VVV light curve database will contain not only periods and magnitudes, but also the variability classes to which the stars are associated~-- which in some cases could/should be confirmed a posteriori using additional data, such as spectra and photometric observations in other bandpasses.

\section{Variability Classes in the VVV Survey: Current Status}\label{sec:classes}

\subsection{Pulsating Variable Stars}\label{sec:puls}

\subsubsection{RR Lyrae and Cepheids}\label{sec:rrlyrceph} 

One of the main scientific goals of the VVV Survey is to complete the census of classical radially pulsating variable stars, such as RR~Lyrae stars and Cepheids, in the Galactic bulge and disk. These stars provide very important means for the study of the 3D structure of the Milky Way because they follow precise PL relations in the near-IR \citep[e.g.,][]{lo86,gbea01,mcea04}, thus they can be used as standard candles, and employed to infer the spatial distribution of their underlying stellar populations. This is particularly efficient if photometric data are available in multiple bands, because in this case the interstellar reddening can be computed on a star-by-star basis, by comparing the measured color index to the intrinsic one, as predicted by the PL relations in different bandpasses. This information can be used both for the distance determinations and for mapping the extinction, and even to trace the large-scale variations in the reddening law in the bulge that has been reported by various studies \citep[see, e.g.,][]{2009ApJ...696.1407N,2013ApJ...769...88N}.

The VVV near-IR data on pulsating stars is particularly powerful in combination with optical (e.g., $I$-band) light curves of complementary time-domain surveys such as OGLE-III, because the accurate mean magnitudes and the large wavelength separation allow a very precise determination of the absolute extinction from the color excess, and the result will be highly independent from the value of the selective-to-total extinction ratio $R_V$. In \citet{idea13b}, we used VVV near-IR photometry of known bulge RRab stars in combination with optical light curves from OGLE-III, to study the 3D structure of the bulge. Our results showed that the spatial distribution of the RR~Lyrae stars is significantly different form the X-shaped distribution of the red clump stars \citep{2010ApJ...724.1491M,2010ApJ...721L..28N,rsea11}; rather, it is rather spheroidal, and does not show a strong bar. This finding implies that the Milky Way may have retained a classical bulge component, with a high fraction of stars in non-cylindrical stellar orbits.

\begin{figure*}
\centering
\plotone{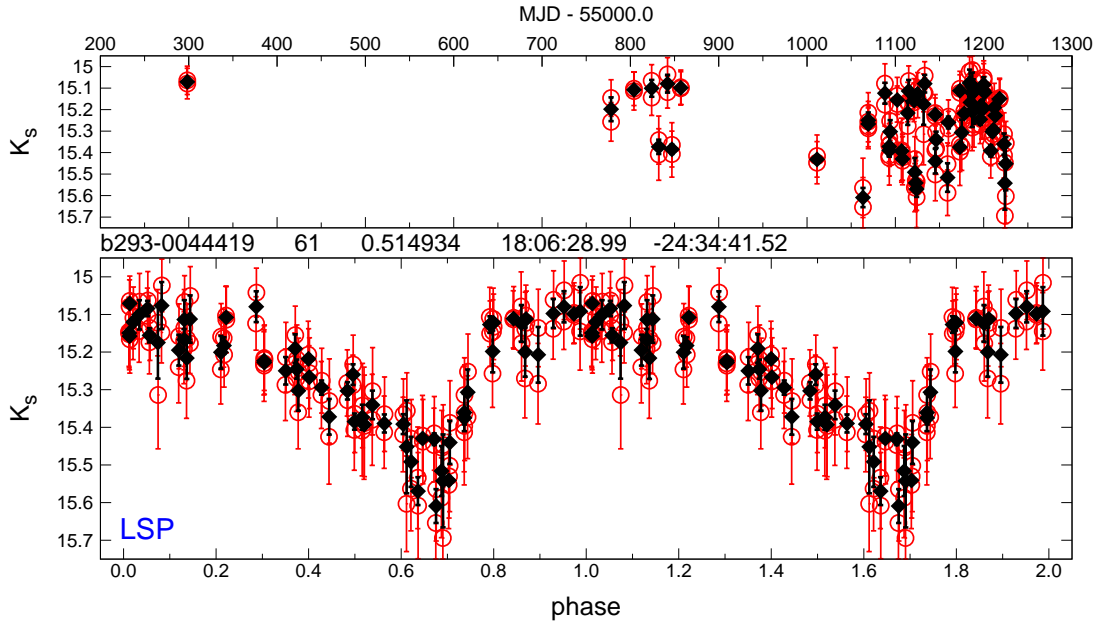}
\vskip0pt
\caption{VDFS 1.2 light curve ({\em top}) and phase diagram ({\em bottom}) for a faint ab-type RR Lyrae star in the bulge, detected by the VVV Survey. The red circles with error bars show the photometric measurements made on {\em pawprints}. The black diamonds with error bars denote the weighted averages of these measurements over tile acquisition sequences, and the corresponding standard error. The star's identifier, the number of $K_s$ epochs, its period (in days) and coordinates ($\alpha$, $\delta$) are shown in the header.}
\label{fig:rrab}
\end{figure*}

The new RR~Lyrae stars discovered by the VVV Survey will allow us to extend our analysis to a much larger sample. Based on our signal detection tests discussed in \S\ref{sec:cross}, we can conclude that the VVV Survey is capable of yielding a highly complete census of fundamental-mode RR~Lyrae stars in the bulge. Figure~\ref{fig:rrab} shows one of the several thousand new RR~Lyrae stars discovered so far by VVV in bulge areas that are not covered by any other time-domain surveys. The high-quality photometry will allow us to unambiguously classify the majority of the new objects, even relatively faint stars, based on their light-curve features (see \S\ref{sec:class}). By greatly expanding the areal coverage as compared to previous optical surveys, and reaching much deeper in highly obscured areas, the VVV Survey will provide unique data that are essential for understanding the structure and formation history of the Milky Way. The same can be expected for classical Cepheids either behind the bulge or in the disk region, since these stars have much larger amplitudes. These objects will provide important means to trace the spiral arm structure on the far side of the Galaxy, which has been out of the reach of optical surveys due to high interstellar extinction, and thus present vast uncharted territories of the Milky Way.

 
\subsubsection{Long-Period Variables}\label{sec:highamp} 
 
As we have seen, one of the main  goals of  the VVV  Survey is to  describe in  detail the
inner Milky Way structure, which is already being accomplished using distance indicators 
such as RR Lyrae stars and red clump giants \citep[e.g.,][]{ogea11,rsea11,idea13b}. However, 
in addition to RR Lyrae and Cepheid variables, long-period variables (LPVs) are also expected
to be detected in large numbers. Since LPVs are very bright and also follow families of PL 
relations \citep[see][for recent reviews and extensive references]{pw12,pw13}, they are 
potentially very powerful distance indicators as well. 

LPV stars, including Miras and semi-regular variables (SRVs), have  been studied  previously in  the
Galactic          bulge           by          different          teams
\citep[e.g.,][]{glass95,kouzuma2009,matsunaga09,isea13b}.  However, these were mostly 
limited to bright objects ($K_s \lesssim 14.3$~mag). VVV represents a
significant  progress in the  search and  study of  LPVs in  the inner
bulge, due  to its  higher spatial  resolution and variability campaign carried 
out in the near-IR,  which allows  one  to  pierce deeply into the  most crowded  and
highly obscured regions  of the Galaxy,  reaching up to $K_s \sim
18.0$~mag in most observed fields \citep{rsea12}.

\begin{figure*}[t]
\begin{center}
  \includegraphics[width=0.85\textwidth]{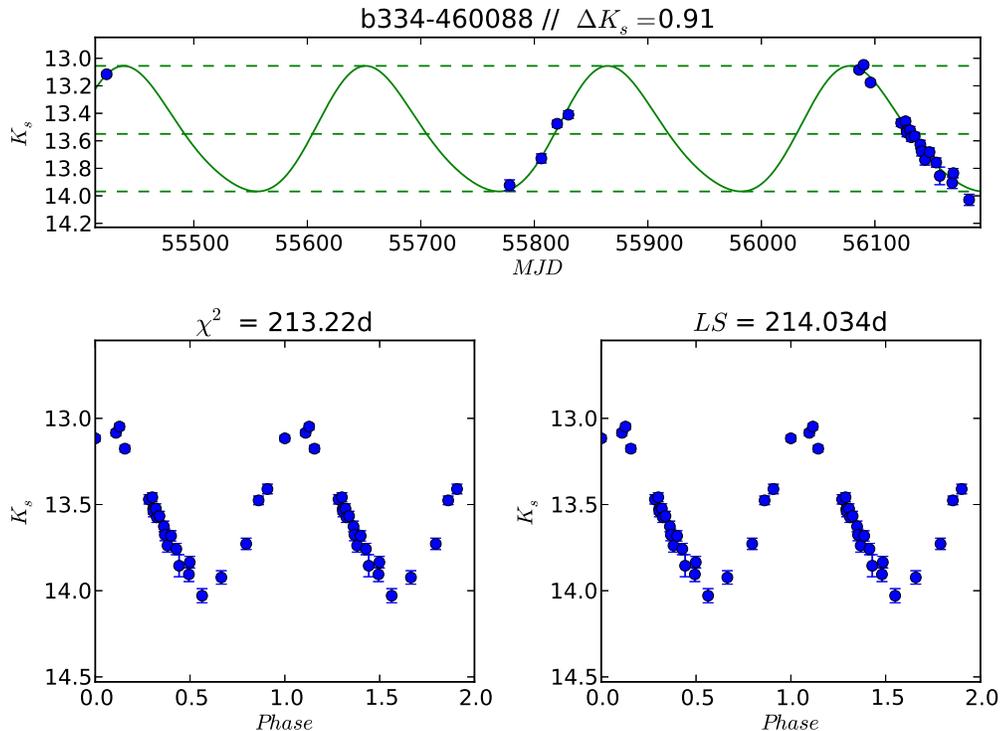}
\end{center}
  \caption{{\em Upper  panel}: VVV  $K_s$-band light curve  of a  Mira
    candidate discovered  in VVV  field b334. The green line represents 
	a Fourier fit, obtained using a period of $P = 213.22$~d, as 
	favored by the Fast $\chi^2$ method \citep{palmer09}. The dashed lines 
    indicated the maximum, average, and minimum light levels, again as 
	obtained from the Fourier fit. 
	{\em Lower left panel}: Folded light curve, using the period favored by the 
	Fast $\chi^2$ method.  
    {\em Lower right panel}: Idem, based on the period favored by the Lomb-Scargle 
	algorithm.}
  \label{fig:lpv}
\end{figure*}

A search for Miras and SRVs in the inner Galactic bulge is already in progress, using 
VVV $K_s$-band data. In this first analysis,  
we selected 12 VVV fields, namely b327 to b338 (see Fig.~\ref{fig:area}),
covering  about  18~square degrees   in the  inner  bulge.   In a  massive
variability survey such as VVV,  Miras  
are relatively easy to find, due 
to the large amplitude of their light curves, which exceed 0.35~mag \citep{sh89,gb05} 
and are on average 0.6~mag \citep{pwea00,mr02} in the $K$-band, 
and periods in the range $100 \lesssim P({\rm d}) \lesssim  1000$. 
About 100 reliable Mira candidates were identified  
from a total of $\sim 17.6$ million light curves  that  were   produced,
with 23-32 data points per object. Each light curve covers the first 
two years  of observations (2010-2012) with different cadence 
depending on the VVV field (see \S\S\ref{sec:status}, \ref{sec:var-over}).

In order  to find the period and  amplitude of each variable,
the light curves of the Mira candidates were fitted by applying both
Fast $\chi^2$  \citep{palmer09}  and   Lomb-Scargle
methods \citep{nl76,js82}.   An  illustrative  example   of  our  results  is  shown  in
Figure~\ref{fig:lpv}; naturally, the derived periods will become better 
defined as more data covering a longer baseline become available. 

A first catalog will be presented in 
Gran et al. (2013, in preparation). In addition to Miras and SRVs, this  
includes a large number of variable AGB stars (previously identified as OH/IR or 
Maser sources, see \citealt{engels05}), besides young stellar objects and other 
still unclassified sources.

\subsection{High-Amplitude Pre-Main Sequence Variables}\label{sec:prems}
 
The majority of pre-main sequence (PMS) stars have measurable variability, which is usually attributable
to magnetic activity. Hot spots formed at the base of accretion funnel flows can also cause variability,
though this would be more important at blue and near-UV wavelengths. Some PMS stars, e.g. KH15D \citep{kh98},
have variable extinction, which can lead to high-amplitude variability even in the IR.
However, the highest amplitudes are observed in eruptive PMS variables, where the mechanism is believed to
be large variations in the accretion rate which directly change the luminosity of the star and the inner
parts of the circumstellar disk. 

Eruptive PMS variables are usually divided into FU Orionis types (FUors) and EX Lupi types (EXors), which 
show occasional variability in excess of 2 to 6 magnitudes on timescales of months (EXors) or years to decades
(FUors; for a review, see \citealt{hk96}). FUors show a rapid rise in luminosity 
followed by a slow decay over decades. Many FUors have associated molecular outflows or jets \citep{ra97,neea94}
in which the effects of historical eruptions on the outflow rate can sometimes be seen \citep{hzea98}.

Fewer than 20 eruptive variables are known in each of the FUor and EXor classes, but it is possible that this
type of variability is common amongst pre-main sequence stars, though intermittent in nature with long intervals
of quiescence. If it is common, this would be important 
for two reasons. Firstly, it may explain the commonly observed scatter in Hertzsprung-Russell diagrams of PMS clusters,
a phenomenon that hampers the assignation of masses to PMS stars with evolutionary tracks, with consequences
for measurements of the initial mass function. Secondly, they may explain the long-standing ``luminosity problem'' 
\citep[e.g.,][]{skea90}, which consists of the fact that low-mass PMS stars are typically less luminous than 
expected for objects that should be above the main sequence while descending a Hayashi track.

As far as eruptive PMS variables go, our work on the VVV data has the potential to (i)~precisely 
quantify the incidence of EXors and (ii)~unveil variability in optically-obscured protostars, an area where
only a small amount of work has been done to date. We might expect the high-amplitude variability to be more common
in these obscured, generally younger systems, in which the average accretion rate is higher. 
We are also beginning to explore the potential of the common low-amplitude variability as a method for tracing 
dispersed PMS stellar associations. This may allow us to investigate the duration of PMS evolution (which will 
be mass-dependent), by comparing the number of variables with the number of stars with disks.

Our first searches for PMS variables were based on VVV Data Release 1 \citep[DR1;][]{rsea12}, which had a typical time baseline of only a few months in 2010 and only 5 epochs of $K_s$ photometry. We searched in the disk region of VVV ($295^{\circ}<\ell<350^{\circ}$, see Fig.~\ref{fig:area}) for candidate variables with $K_{s,{\rm max}}-K_{s,{\rm min}}>1$~mag, which is the approximate upper limit for magnetic variability. We also did a more general search for variables with RMS $K_s$ variability $>0.2$~mag, in order to see whether that would pick out star formation regions (SFRs). Both searches of the catalogs produced a large number of false positive candidates requiring visual inspection, but their number was reduced by requiring a detection in the {\it Spitzer} GLIMPSE Legacy survey \citep{rbea03}, and in particular by focusing on the subset of 18,949 very red sources ($[4.5]-[8.0]>1$~mag) identified in GLIMPSE by \citet{trea08}. Most of the Robitaille sources are candidate protostars. The $K_{s,{\rm max}}-K_{s,{\rm min}}>1$~mag search identified 1881 candidates in VVV, of which 47 are detected in GLIMPSE and 12 were known as red Robitaille sources. These 12 were all confirmed as genuine variables by inspection of the images. They have $K_{s,{\rm max}}-K_{s,{\rm min}}$ between 1.02 and 1.76 mag, and the majority are undetected in the $J$ and $H$ passbands, although they are well above the $K_s$ detection limit. 

\begin{figure*}
\centering
\plotone{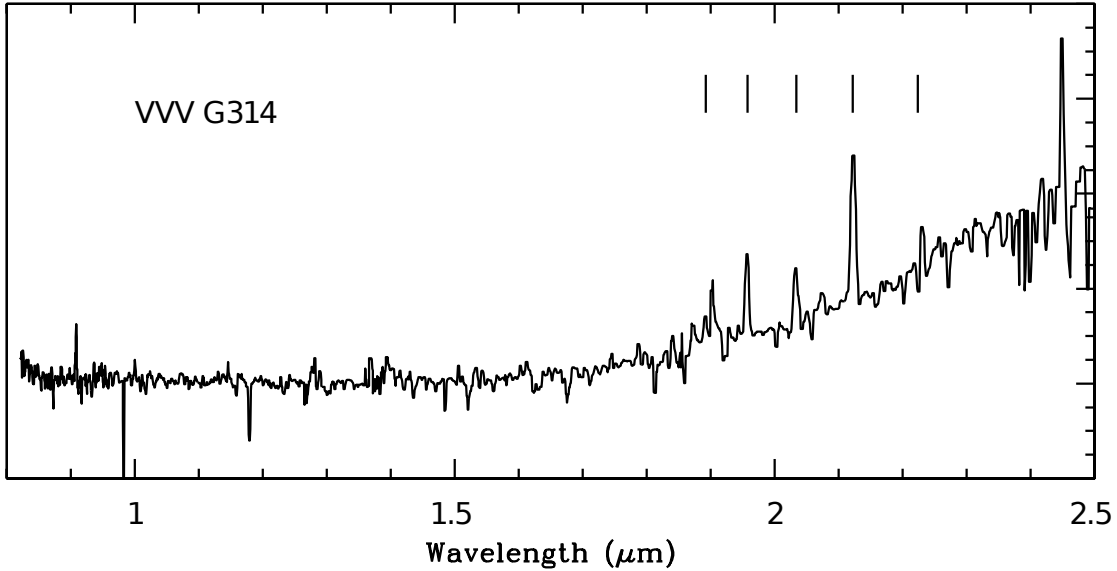}
\vskip0pt
\caption{Near-IR Magellan/FIRE spectrum of G314, a high-amplitude VVV variable that is also in the list of 
\citet{trea08} red {\em Spitzer} GLIMPSE sources. Essentially no flux is seen at $\lambda<1.7 \, \mu$m, and several 
strong H$_2$ emission lines are detected, marked with vertical lines. This is a preliminary reduction of the data.}
\label{fig:g314}
\end{figure*}

A low-resolution spectrum was obtained for one of these red variable sources (G314, see Fig.~\ref{fig:g314}) 
with Magellan/FIRE \citep{rsea08} in March 2012. The spectrum in Figure~\ref{fig:g314} is based on a preliminary reduction, but it clearly shows several very strong emission lines of H$_2$, which are marked with vertical lines. Moreover, the emission was spatially extended by a few arcseconds along the slit, demonstrating the presence of a large-scale outflow.

\begin{figure*}
\centering
\plottwo{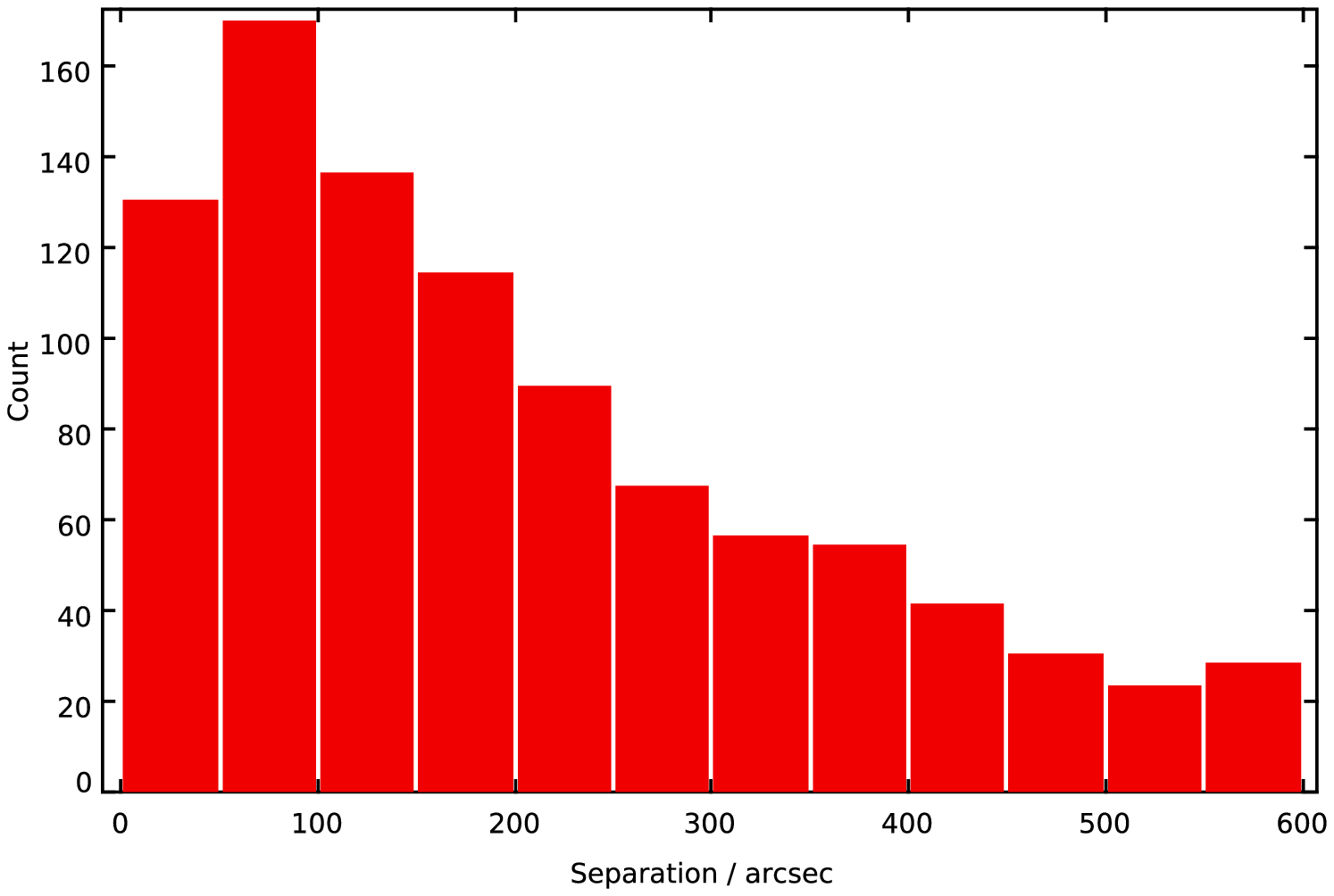}{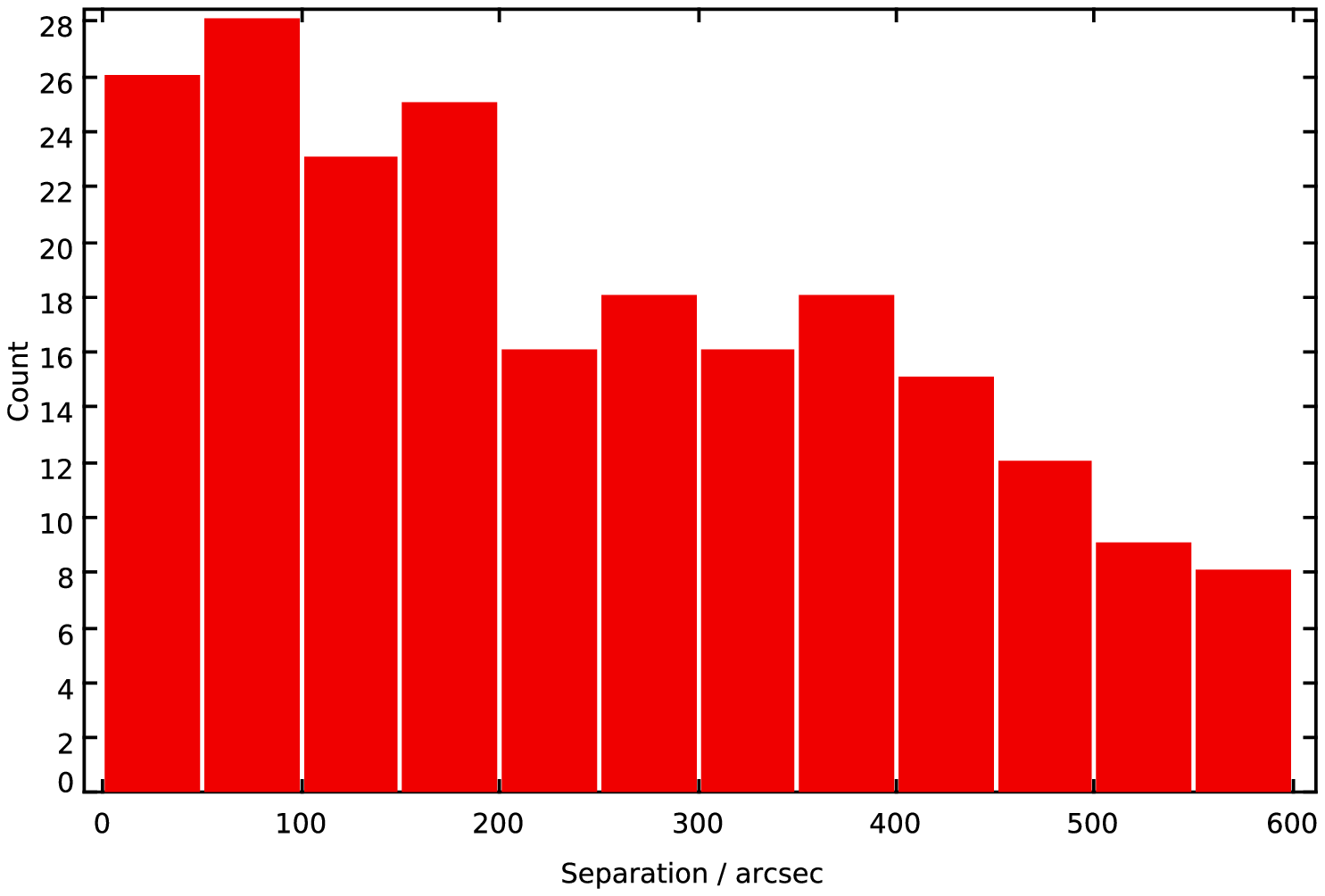}
\vskip0pt
\caption{({\em left}) The distribution of separations of candidate VVV variables from SFRs. The separations are given in units of arcseconds. The rise
to smaller separations demonstrates that these are not chance associations. ({\em right}) A similar histogram for the subset
of candidate VVV variables that have counterparts in the {\it Spitzer} GLIMPSE catalog.}
\label{fig:sfr}
\end{figure*}

An additional search was made for lower-amplitude variables with RMS magnitude variations $>0.2$~mag in the DR1
data. This returned 24,798 candidates, of which 703 are in GLIMPSE and 41 objects are Robitaille red sources 
(including the 12 with $K_{s,{\rm max}}-K_{s,{\rm min}}>1$~mag). We cross-matched these candidate variables which have {\it not} been visually
inspected with the \citet{va02} catalog of SFRs to see whether there was a significant spatial 
association. We found that 4\% (937/24,798) of the candidate variables are within 10 arcminutes of a known SFR and the histogram of candidate to SFR separations shows a rising number of matches with decreasing 
separation. This indicates a clear asociation of at least a significant minority of VVV variables with SFRs 
(see Fig.~\ref{fig:sfr}). Similarly, we found that 30\% (214/703) of the GLIMPSE sources are within 10 arcminutes of a known SFR, 
and the separation histogram has a similar trend. To establish the true fraction of VVV variables that are PMS stars
will require visual inspection of a large sample and careful checking of the photometry to avoid the errors that can
arise in crowded fields.

Subsequently, we undertook a search for very high-amplitude variables in the combined 2010 to 2012 datasets for the
disk region, focusing on the region at $-1^{\circ}<b<1^{\circ}$ 
(Contreras Pe\~na et al. 2013, in preparation). Following visual inspection of a 
few hundred candidates, a total of 77 genuine variables were identified with $K_{s,{\rm max}}-K_{s,{\rm min}}>2$~mag, which is the level 
more typically associated with eruptive variability when long time baselines 
are available. Spectroscopy from 0.8 to 2.5~$\mu$m has now been obtained for 6 of these systems in April 2013, using 
Magellan/FIRE in echelle mode at $R=6000$. Priority was given to sources (i)~with a likely association with clusters or 
SFRs of known distance; (ii)~with highest amplitude; (iii)~with signs of a recent outburst in the light curve; 
(iv)~bright enough for quick observation. A further 4 systems with slightly lower amplitude in the G305 star forming
complex were also observed, and G314 was re-observed. A few of these 11 systems also showed strong, spatially extended 
H$_2$ emission, similar to that seen in G314. Data reduction and analysis is currently ongoing.

\subsection{Eclipsing Variable Stars}\label{sec:eclips}

\begin{figure*}
\centering
\includegraphics[width=0.75\textwidth]{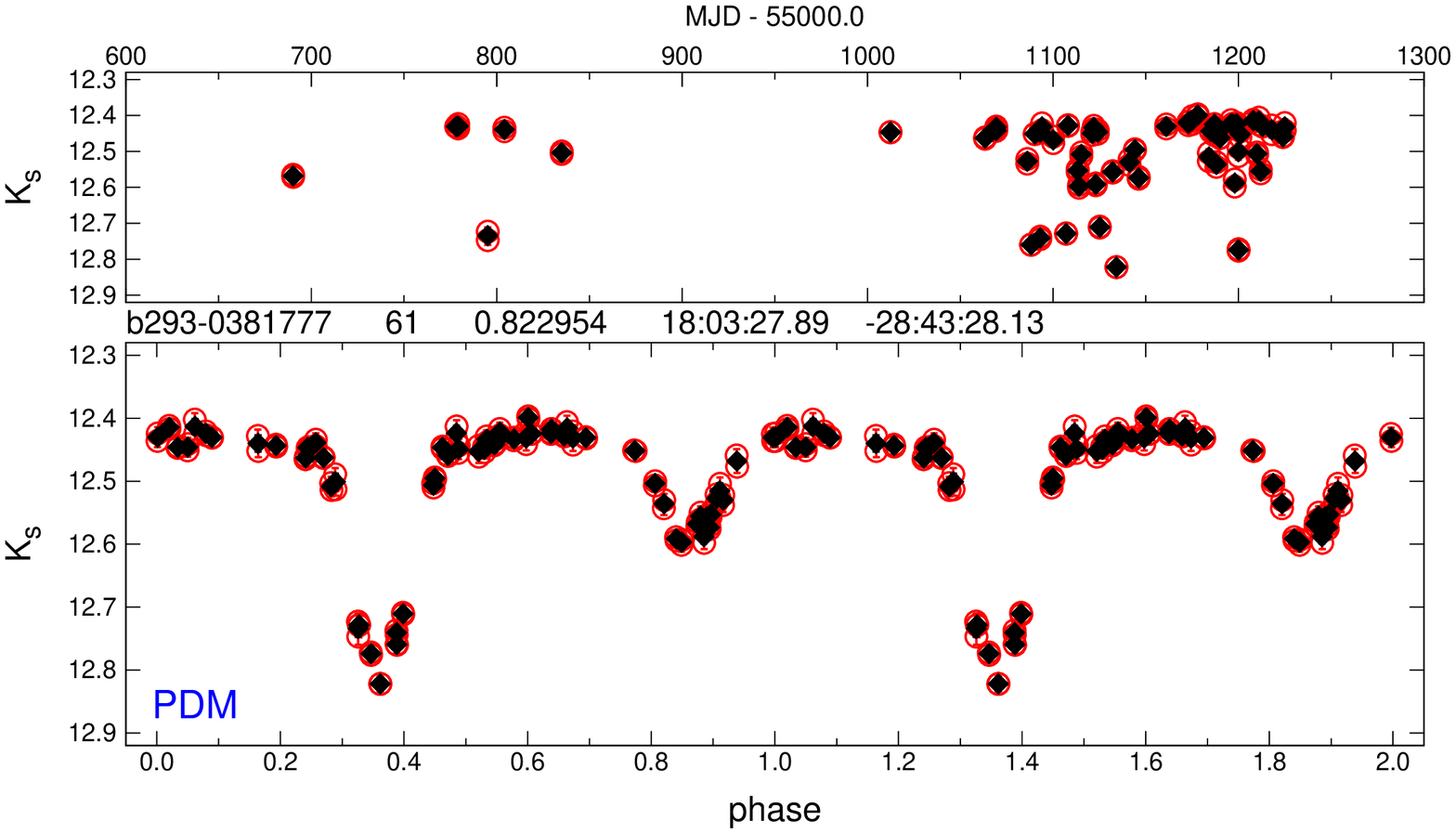}
\vskip6pt
\includegraphics[width=0.75\textwidth]{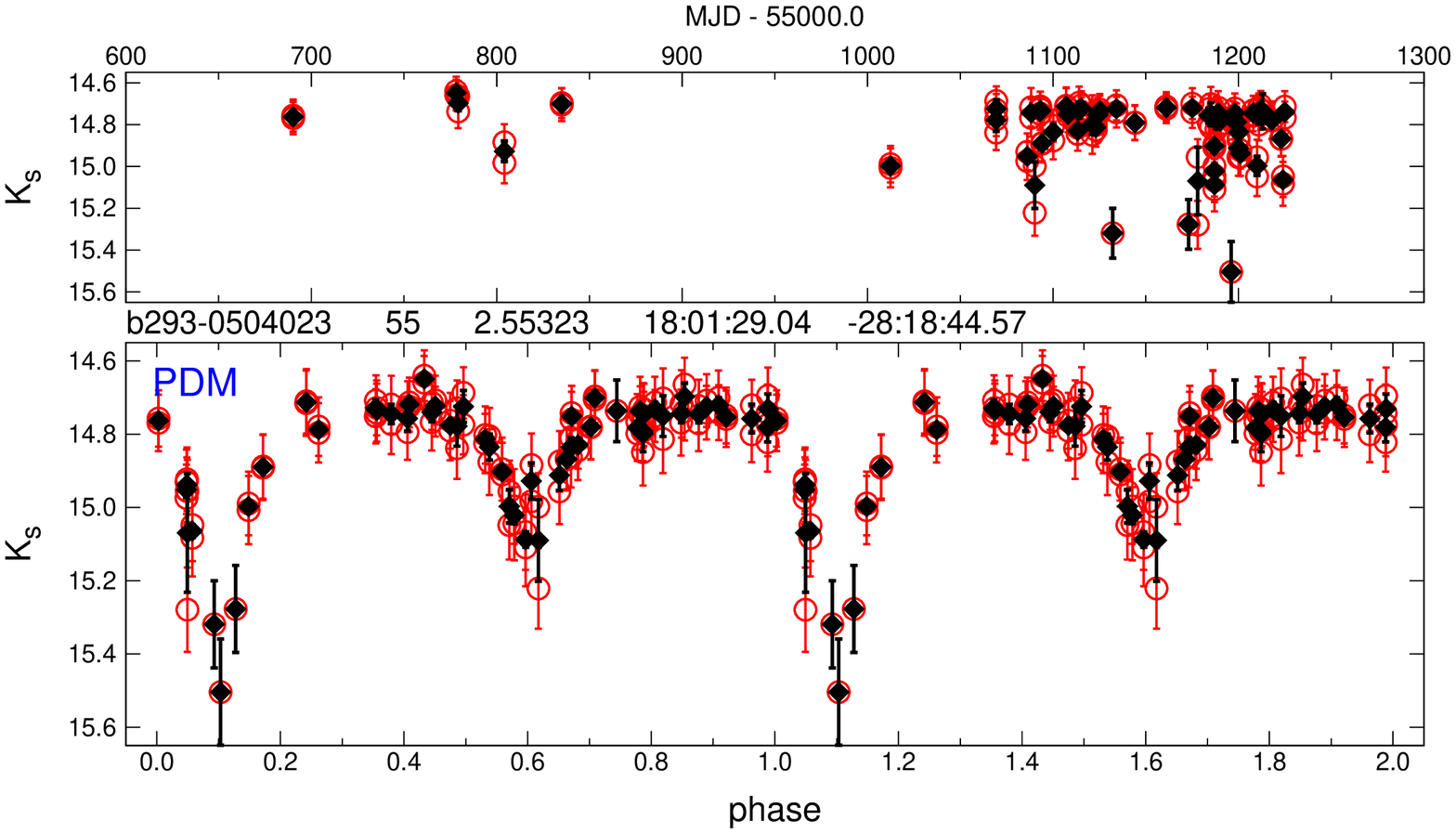}
\vskip0pt
\caption{As in Figure~\ref{fig:rrab}, but showing two detached eclipsing binary systems in the bulge, as detected by the VVV Survey.}
\label{fig:ebs}
\end{figure*}

Eclipsing binary systems in general, but detached systems especially, are extremely important objects in astrophysics, for they provide one of the most robust means of deriving stellar radii, masses, and even ages \citep[e.g.,][]{dp80,ja91,bp96}, thus providing a ``royal road'' to stellar astrophysics \citep[][and references therein]{ab05,js12}. The VVV Survey will provide a huge number of eclipsing systems of all types, and we have already started to travel along this ``royal road'' using VVV data. Examples of VVV light curves for two eclipsing binary systems can be found in Figure~\ref{fig:ebs}. 

As an example of their use in astrophysics, a method to use the VVV data for tracing the structure of the Milky Way with detached eclipsing binaries (DEBs) was  introduced in \citet{khea13}. The idea behind the method is that one can calculate the distance by comparing the derived absolute magnitudes from the model with the observed ones. This normally requires spectroscopic observations to calculate radial velocities of both components, used for calculating the masses and orbital parameters, crucial for obtaining the true absolute values of the stellar parameters (like the radii). One also needs accurate theoretical atmosphere models and/or calibrations to infer surface brightness on the basis of colors, line ratios or other observables \citep{pac97}. 

With the increasing availability of extensive databases of state-of-the-art stellar evolution models, it is now possible to derive a complete set of physical parameters of a binary's components without time-consuming spectroscopy, from their light curves only. In our approach we implemented two codes, which can analyze large numbers of DEB light curves in a short time. The first one~-- the Detached Eclipsing Binary Light curve fitter (DEBiL)~-- is a program which rapidly fits DEB light curves to a simple, geometric model \citep{dev04,dev05}. The second code~-- Method for Eclipsing Component Identification \citep[MECI;][]{dev06a,dev06b}~-- fits a physical model to each DEB using readily available photometric data only. It is designed to work from the DEBiL model as a starting point, building an improved physical model of the DEB therefrom. MECI assumes that the binary's stellar components formed together and evolved along their respective evolutionary tracks, without any mass transfer. If observed magnitudes in different bands are given, MECI returns the absolute magnitudes, so the distance may be calculated directly. Finally, to deal with the strong and variable interstellar reddening, we implemented the idea of reddening-free indices, introduced in \citet{mcea11}. These indices are combinations of magnitudes in 3 bands, in the form of

\begin{equation}
m_X = m_1 - c(m_2 - m_3), 
\label{eq:mx}
\end{equation}

\noindent where $m_{1,2,3}$ are the apparent magnitudes in the available bands, 
and $c$ is a multiplication coefficient dependent on the extinction law assumed. 
The coefficient $c$ is given in such way that, within a given extinction law,
defined in terms of the ratios of extinction values in the given bands 
$A_1 : A_2 : A_3$, the following equation is true:

\begin{equation}
m_X = m_{X,0} = m_{1,0} - c(m_{2,0} - m_{3,0}), 
\label{eq:mx0}
\end{equation}

\noindent where the ``0'' subscripts indicate quantities corrected for reddening, 
and thus $m_1 = m_{1,0} + A_1$, $m_2 = m_{2,0} + A_2$, $m_3 = m_{3,0} + A_3$. Combining 
equations~(\ref{eq:mx}) and (\ref{eq:mx0}), we have: 

\begin{equation}
 m_{1,0} + A_1 - c \left[(m_{2,0} + A_2) - (m_{3,0} + A_3)\right] = m_{1,0} - c(m_{2,0} - m_{3,0}). 
\label{eq:mxd1}
\end{equation}

\noindent Thus,  

\begin{equation}
 c = \frac{A_1}{A_2 - A_3}. 
\label{eq:cdef}
\end{equation}

Naturally, the definition in equation~(\ref{eq:mx}) is also applicable to {\em absolute} 
magnitudes, and so 

\begin{equation}
M_X = M_1 - c(M_2 - M_3), 
\label{eq:mabsx}
\end{equation}

\noindent where the capitalized $M$'s denote absolute magnitudes. On this basis, we obtain, for the apparent 
distance modulus, 

\begin{equation}
 m_X-M_X = (m-M)_X = \left[m_1 - c(m_2 - m_3)\right] - \left[M_1 - c(M_2 - M_3)\right], 
 \label{eq:mX-MX}
\end{equation} 

\noindent which implies 

\begin{equation}
 (m-M)_X = (m_{1,0}+A_1) - c \left[(m_{2,0}+A_2) - (m_{3,0}+A_3)\right] - \left[M_1 - c(M_2 - M_3)\right].  
 \label{eq:mX-MXd2}
\end{equation} 

\noindent This can be rewritten as 

\begin{equation}
 (m-M)_X = \left[m_{1,0} - c (m_{2,0}-m_{3,0})\right] - \left[M_1 - c(M_2 - M_3)\right] + \left[A_1 - c  (A_2 - A_3)\right].  
 \label{eq:mX-MXd3}
\end{equation} 

\noindent Thus,   

\begin{equation}
 (m-M)_X = (m-M)_{0} -  c (m-M)_{0} + c (m-M)_{0} + \left[A_1 - c  (A_2 - A_3)\right], 
 \label{eq:mX-MXd4}
\end{equation} 

\noindent and so, using equation~(\ref{eq:cdef}),  

\begin{equation}
 (m-M)_X = (m-M)_0 = 5 \, (\log d - 1), 
 \label{eq:distmod}
\end{equation} 

\noindent where $d$ is the distance in pc. Therefore, {\em reddening-free magnitudes are ideally suited for the calculation of distances, without the need for prior knowledge of extinction values towards individual stars}. 

The VVV variability campaign is still ongoing (\S\ref{sec:var-over}), and a catalog of eclipsing binaries has thus not yet been prepared. Therefore, we used the eclipsing binaries from the OGLE-II variable stars catalog \citep{woz02}, for which $I$-band light curves are available. \citet{dev05} presented models for about 10,000 of them, obtained with the DEBiL code, and identified 3170 of them as ``detached.'' To build the reddening-free indices we used $J,H,K_s$ photometry from VVV, in addition to $V,I$ data from OGLE-II, and assumed a canonical extinction law with $R = 3.09$. For consistency, we also built a set of reddening-free isochrones, based on the Padova models \citep{gir00,mar08}.

\begin{figure*}
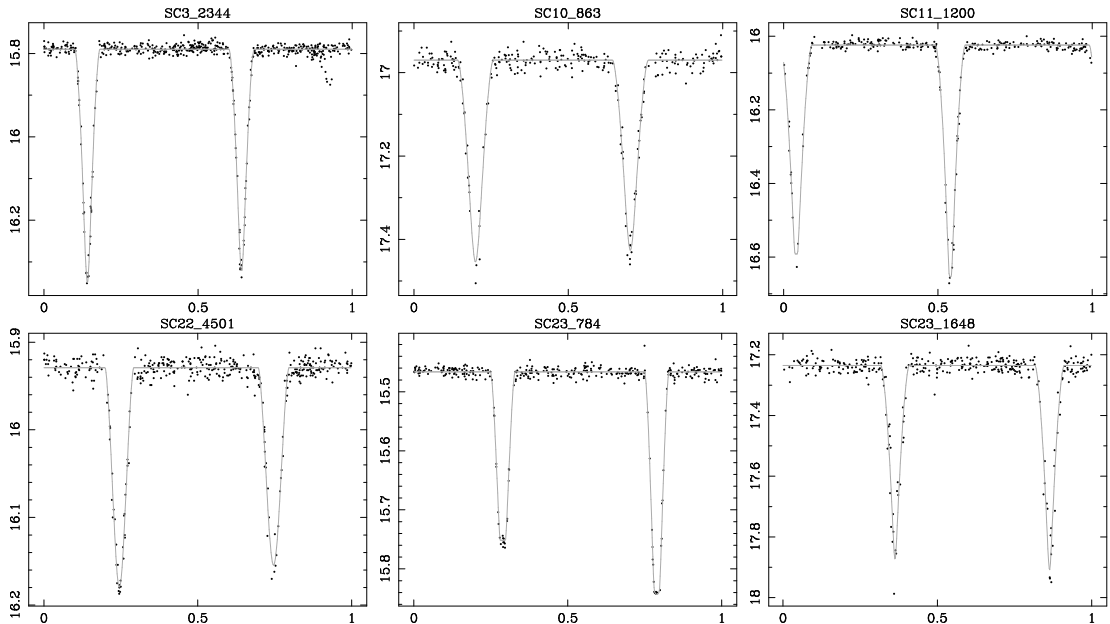

\centering
\includegraphics[width=0.31\textwidth]{lc_sc3_2344.eps}
\includegraphics[width=0.31\textwidth]{lc_sc10_863.eps}
\includegraphics[width=0.31\textwidth]{lc_sc11_1200.eps}
\includegraphics[width=0.31\textwidth]{lc_sc22_4501.eps}
\includegraphics[width=0.31\textwidth]{lc_sc23_784.eps}
\includegraphics[width=0.31\textwidth]{lc_sc23_1648.eps}
\caption{OGLE $I$-band light curves (dots) and MECI models (lines) for a sample of the researched systems.}\label{fig_2.4_lc}
\end{figure*}

\begin{figure*}
\centering
\includegraphics[scale=0.6, angle=90]{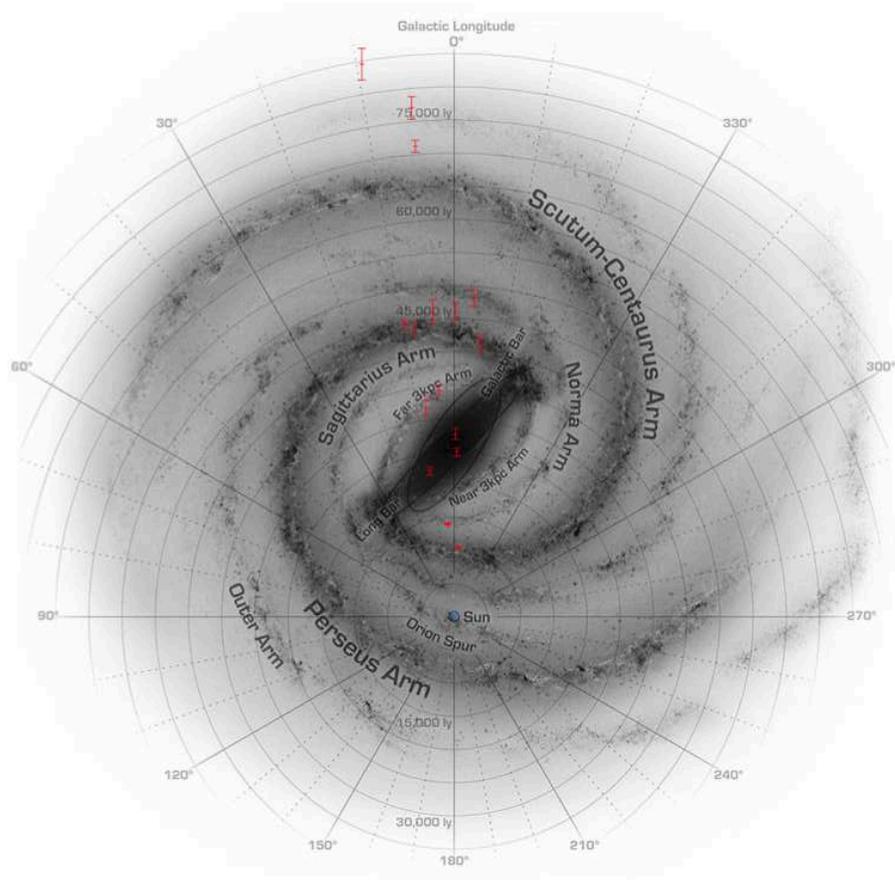}
\vskip0pt
\caption{Positions of the sixteen closest eclipsing binaries in our study (red points with error bars), plotted over a reconstruction of the Milky Way plane from \citet{chu09}. The position of the Sun is shown as a blue asterisk. Several spiral disks are labelled according to their names. 
}
\label{fig_2.4_rec}
\end{figure*}

The procedure just described successfully yielded physical parameters, ages, and distances to 23 DEB systems. In Figure~\ref{fig_2.4_lc} we present examples of the utilized OGLE-II light curves, superimposed on the best-fitting MECI models. In Figure~\ref{fig_2.4_rec} we present the positions of the 16 closest systems in the Galactic plane $(X,Y)$, with the Sun located at (0,0) and the Galactic center at (8,0)~kpc. They are plotted over a reconstruction of the Milky Way from \citet{chu09}. 

The locations of at least 13 of the studied targets coincide well with major structures of the Milky Way, such as the bulge and spiral arms, including Scutum-Centaurus, Norma, Far 3~kpc, and Perseus. This shows that our approach is suitable for tracing the structure of the Milky Way. It is notable how many objects were found in a poorly studied area behind the bulge, which proves that the combination of near-IR VVV data with optical photometry from other sources can be a powerful tool for studying this part of the Milky Way. Another 7 systems (not shown on the reconstruction) were found at larger distances, corresponding to the Sagittarius stream~-- a structure of stars related to the Sgr dSph galaxy \citep[e.g.,][]{skea13}. These Sagittarius DEB candidates, the first to be identified in that galaxy's stream, will be the subject of detailed spectroscopic follow-up by our team.

\subsection{Cataclysmic Variable Stars}\label{sec:cvs}
Cataclysmic  variable stars (CVs), including novae and dwarf novae, consist of tight binary systems in which one of the components~-- the so-called {\em primary}~-- is a WD star, and the other component~-- the {\em secondary}~-- is a low-mass, K- or M-dwarf (i.e., main sequence) star. In CV systems, the dwarf star has filled its Roche lobe, and is transferring matter onto the primary. CVs are thus eruptive binary systems, with the outbursts being detectable over a wide wavelength range, including the near-IR. However, the quiescent light curves of 
dwarf novae  and related objects  are dominated by the  orbital motion, 
and  their behavior in  the near-IR  differs from  that seen  at optical
wavelengths.  While in the optical  the variations are produced by the
hotter  parts of  the  system,  namely the  primary  star and/or  the
accretion disk,  in the  near-IR the light curve  is dominated  by the
emission of the cool, late-type  secondary star, whose spectra peak at
longer wavelengths \citep[e.g.,][]{ch01}.  In close binaries the
secondary star  will be  distorted by tidal  effects when close  to or
filling its  Roche lobe.  The  orbital motion thus  causes ellipsoidal
variations which are more prominent in near-IR light curves.  This
modulation enables us to determine the orbital parameters of the system
and even  to map  the surface brightness  distribution of  these stars
\citep[e.g.,][]{trea07,trea10}.


The recent discovery of dozens of new dwarf novae CVs by the OGLE team
demonstrates the ability  of large surveys to search  for new variable
sources   even   in  the   most   crowded   regions   of  the   Galaxy
\citep{pmea13}. Long-period symbiotic systems, in which the 
secondary star is a late-type giant rather than a main-sequence star, 
have also been discovered in both the OGLE and MACHO data 
\citep{bmea13}. 
In like vein, we expect large numbers of interacting binary systems to 
show up in the VVV data. Indeed, in the case  of novae,  preliminary results  
have already been presented  in \citet{rsea13}.

%

There are  about 400 known novae in  the Galaxy, with  $\sim35\%$ of them falling in 
the VVV Survey area.  Interestingly, the spatial distribution of novae shows
a  ``zone  of avoidance,''  with  just a  few  objects  belonging to  the
innermost  regions.  Moreover,  the comparison  with  nearby galaxies
suggests that  we lose  many nova eruptions  every year 
\citep[][and references therein]{rsea13}. 
Not  surprisingly, the regions  with a
lack  of   objects  are  the regions most heavily obscured by dust,   beyond  the
capabilities  of the  current searches  for novae, which are mostly carried out in  
the optical.  \citet{rsea13} provide  $JHK_s$ data for  93
Galactic novae.   For some of these objects,  colors have been  reported for
the first  time in a homogeneous  dataset, since a  large fraction of these 
novae  were beyond  the detection  limit of  previous  near-IR surveys
\citep[e.g., 2MASS;][]{msea06}.

\begin{figure*}[ht!]
\begin{center}
  \includegraphics[height=0.75\textheight]{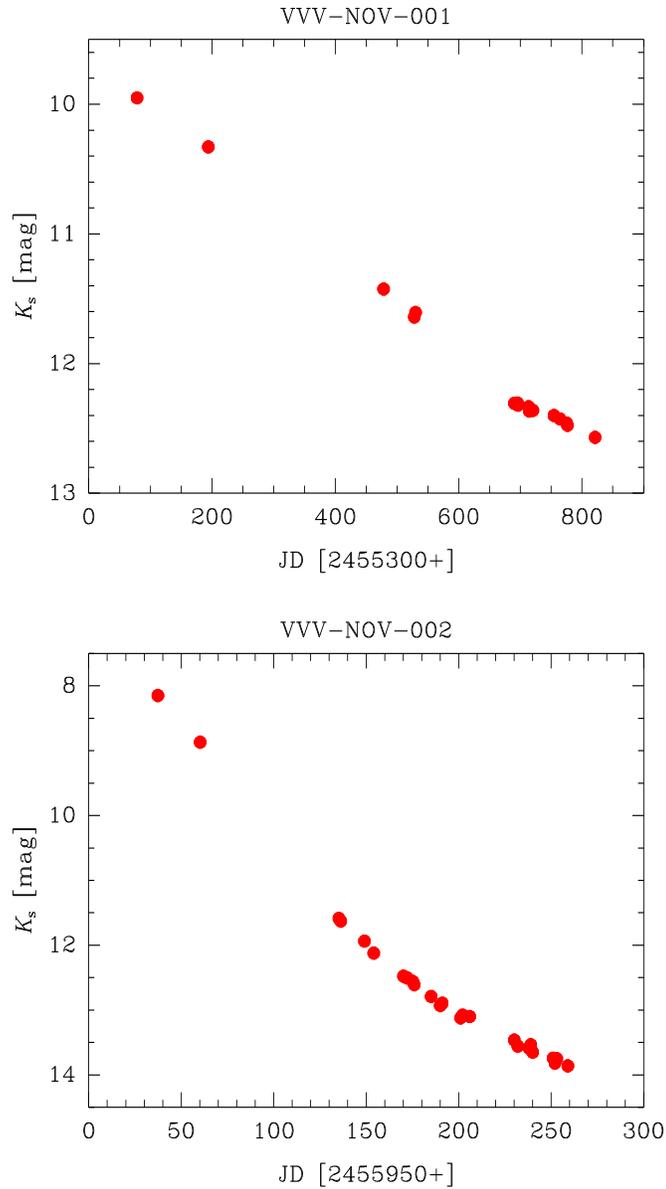}
\end{center}
  \caption{VVV $K_{\rm s}$-band light curves of VVV-NOV-001 ({\em upper panel}) and VVV-NOV-002 ({\em lower panel}).}
  \label{fig:lcurve}
\end{figure*}

The low novae  rate observed in the Galaxy encouraged  us to start a
search  for the hidden  novae in  the aforementioned ``zone  of avoidance''  region. We
thus selected  24 VVV  fields covering  about 36~square degrees   towards  the inner
Galactic  bulge.  In this  first attempt,  we made  use of  the first
variability tables  available, limited to $20-40$~epochs  in the inner
bulge at the  time of writing.  Even with a  limited number of epochs,
the  VVV data allowed   us    to   discover   two  new  Galactic   novae   candidates
\citep[VVV-NOV-001                                                  and
VVV-NOV-002;][]{rsea12a,rsea13b},  and a few more objects are currently 
under study. VVV light curves of VVV-NOV-001 and VVV-NOV-002
are  presented  in  Figure~\ref{fig:lcurve}.   Spectra taken  with  the SOAR
telescope  show the  presence of  emission lines  which are typical of  novae in
VVV-NOV-001 (Saito et  al.  2013c, in preparation), while  optical data from
OGLE     confirmed      VVV-NOV-002     as     a      D-class     nova
\citep{wu13}.  While VVV-NOV-001  was discovered on the
Galactic plane, with $(\ell,b)=(8.89,-0.16)$~deg,  VVV-NOV-002 is one of the
closest novae to the Galactic center known, with $(\ell,b)= (-2.28,1.97)$~deg.
A search for  novae will be extended to  the high-extinction regions
of the inner disk in the near future.

\subsection{Microlensing Events}\label{sec:microl}
Gravitational microlensing events can be of enormous astrophysical significance \citep[e.g.,][and references therein]{bp96a,sm12}. 
Models of the spatial dependence of the microlensing optical depth $\tau$ \citep{ekea09} show that IR surveys like VVV can be very efficient in the search for microlensing events, and can probe directly the mass distribution contained in the inner regions of the Galaxy. Unfortunately, existing optical microlensing searches based on CCD detectors do not cover the whole bulge or the plane (although OGLE-IV\footnote{\tt http://ogle.astrouw.edu.pl} already represents an important step forward in this regard), and in particular, they miss the inner regions where this optical depth is higher, thus poorly constraining the models. Therefore, one of the main goals of the VVV Survey is to search for microlensing events in the inner Milky Way. We are especially interested in looking for rare events in our data, such as: 

\begin{figure*}
\centering
\includegraphics[width=0.95\textwidth]{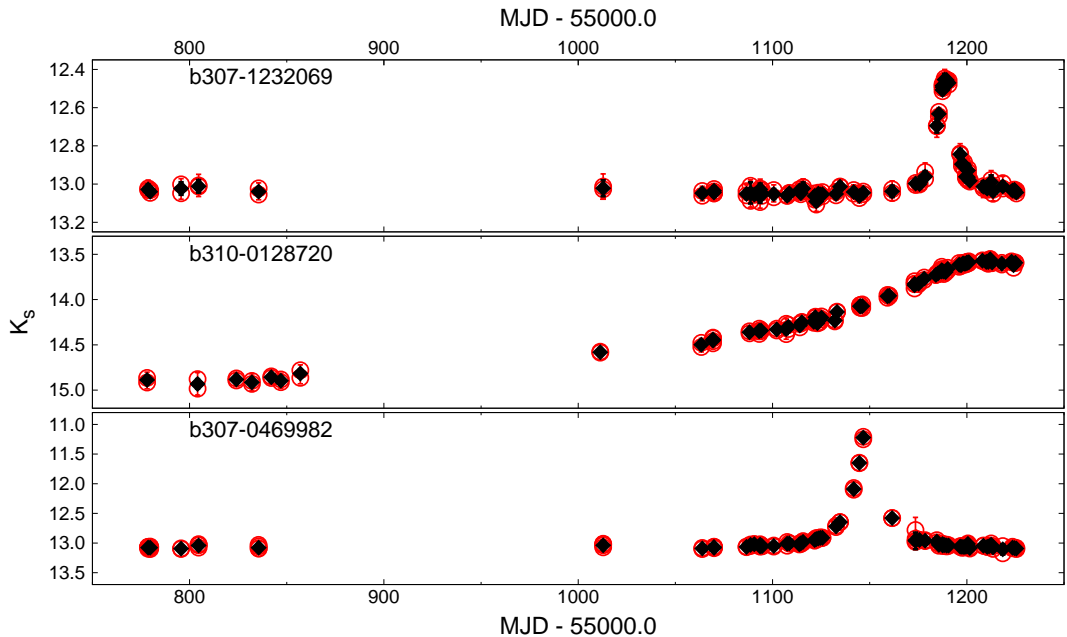}
\vskip0pt
\caption{Three microlensing events, as detected in the course of the VVV Survey.}
\label{fig:ml}
\end{figure*}

\begin{itemize}
\item Very reddened events; 
\item Short-timescale events, due to planetary or brown dwarf microlensing. The advantage of using microlensing to search for planets is that there is no preference for nearby objects or bright stars, contrary to what occurs with other techniques. Thus, microlensing allows us to further probe the planet parameter space, searching, for instance, for planets with periods that are too long to be detected by other techniques, or not sufficiently close to the star as to produce detectable Doppler shifts in their spectra \citep{md00};  
\item Very long-timescale events, due to massive black holes. Such a long-term variability study is possible with our $\approx 7$-year-long survey, adequately covering the baseline for these long-timescale microlensing events; 
\item Binary microlensing events;  
\item Parallax events; 
\item High-magnification events in obscured dense fields;  
\item Microlensing of source stars in the Sgr dSph galaxy \citep[e.g.,][]{ppea05}.
\end{itemize}

\begin{figure*}
\centering
\includegraphics[width=0.72\textwidth]{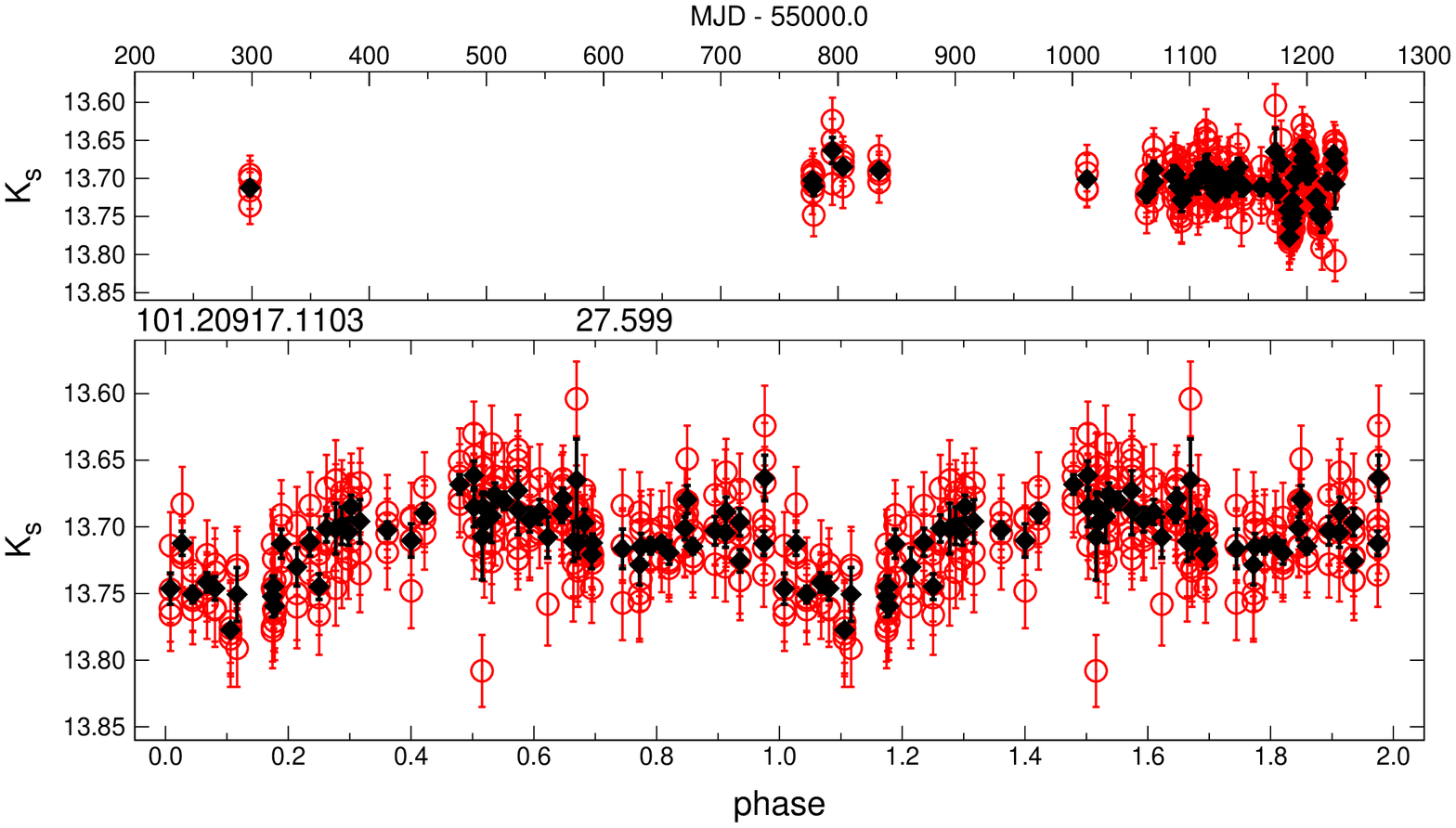}
\includegraphics[width=0.72\textwidth]{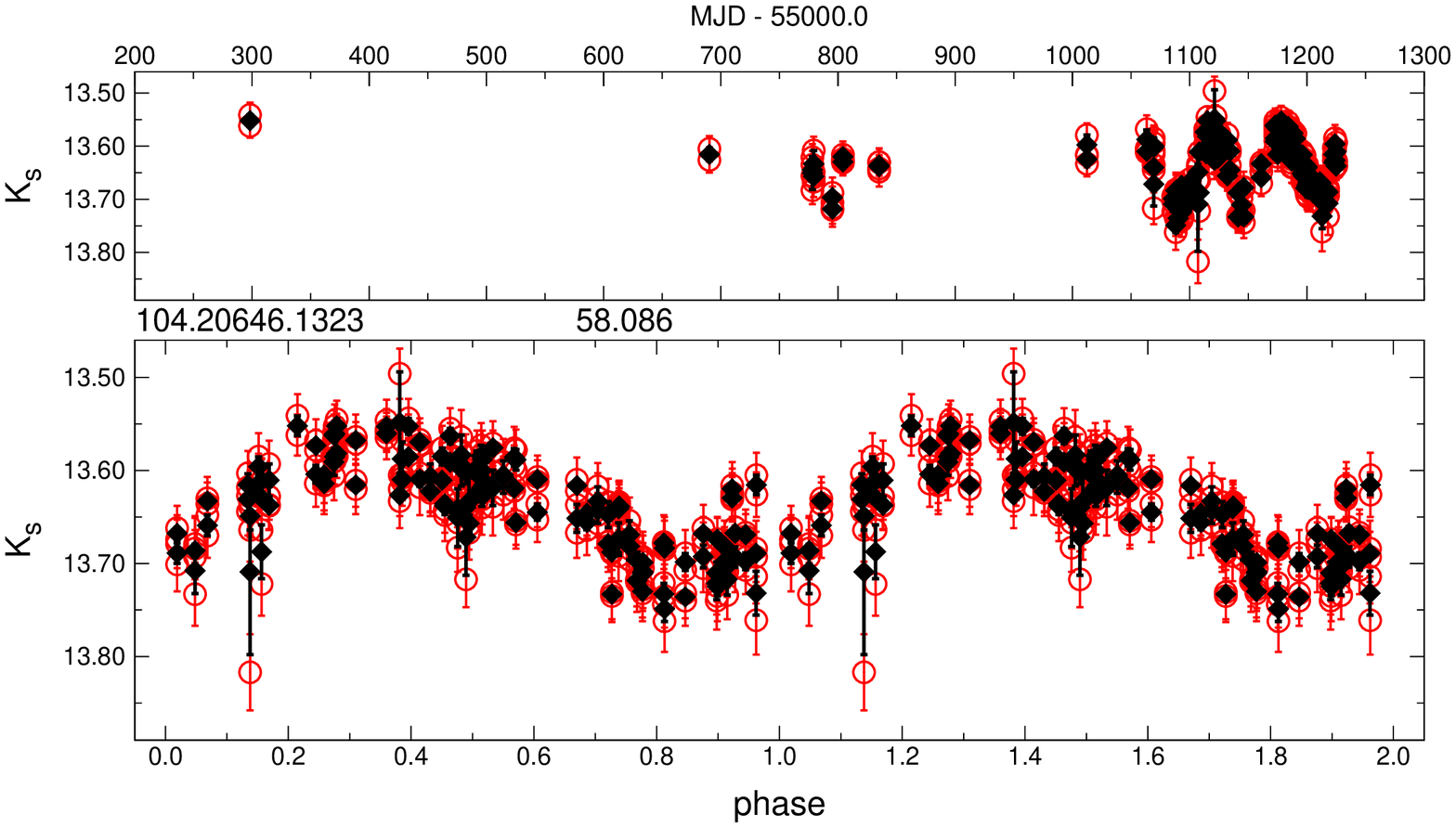}
\includegraphics[width=0.72\textwidth]{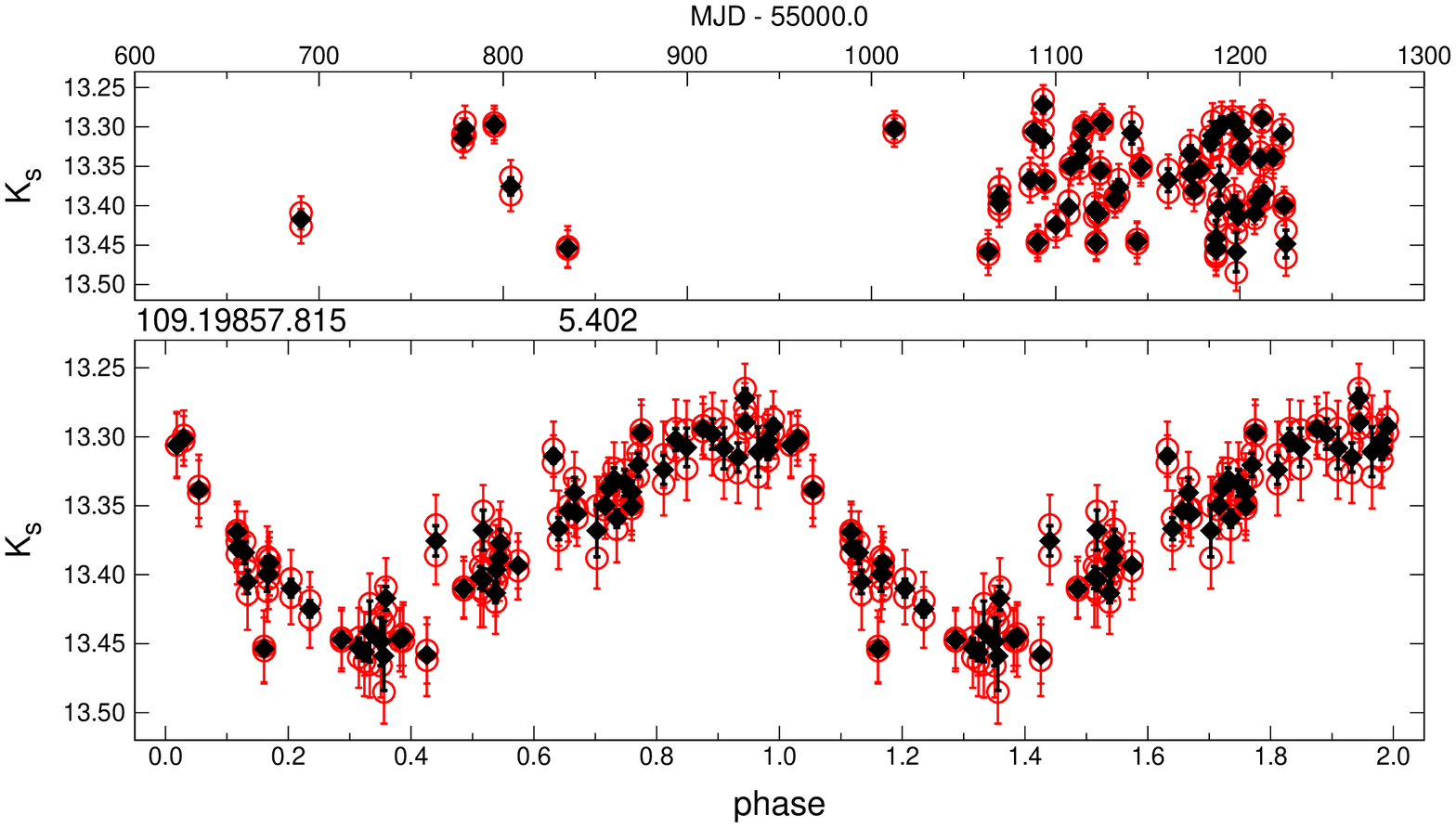}
\vskip0pt
\caption{As in Figure~\ref{fig:rrab}, but showing light curves of stars classified as RS CVn systems by \citet{ajd06}.}
\label{fig:rscvn1}
\end{figure*}

So far we have concentrated our searches mostly on the regions that have not been previously searched for microlensing. These fields are located in the innermost regions of the bulge, where the microlensing optical depth is expected to be high. A quick search has so far revealed about two dozen bulge microlensing events with high amplification. The light curves and the parameters fitted show a range of timescales from a few days to several months, consistent with previous (complementary) results from the OGLE survey \citep{au00}. Some examples of microlensing events are provided in Figure~\ref{fig:ml}.

\subsection{Rotating Variables}\label{sec:rot} 
The Galactic bulge is known to contain a significant number of chromospherically active and rotating variable stars, including stars showing ellipsoidal light curve modulation, FK Com stars, and RS CVn systems \citep{ajd06}. We have cross-matched the \citeauthor{ajd06} catalog, which is based on MACHO observations, with the VVV photometric catalogs as provided by the CASU VDFS 1.2 pipeline, successfully obtaining $K_s$-band light curves for numerous stars previously classified as rotating variables. Some examples of the corresponding VVV light curves are provided in Figure~\ref{fig:rscvn1}. These light curves confirm that VVV near-IR data can be successfully used even to study systems whose light curve amplitude does not exceed $\approx 0.1$~mag in $K_s$, thus also helping properly classify objects whose previous classification may be unclear or ambiguous.

\subsection{Variable Stars in Star Clusters}\label{sec:clusters} 
A significant number of Galactic star clusters fall inside the area surveyed by the VVV: 36 known globular clusters (GCs) and 355 known open clusters \citep[OCs;][]{dmea10}, in addition to newly discovered GC \citep{dmea11,mo11} and OC \citep{bo11,ancea12,ancea13} candidates. The inner Galactic star systems are generally deeply buried behind a curtain of dust and gas that hides them in optical observations, and so our knowledge of their variable star populations in particular is at present very incomplete \citep[e.g.,][]{mcea06}. Fortunately, the highly diminished extinction in $K_s$ ($A_{K_s}\sim0.1 \, A_V$) will help us unveil the variable stars that are present in these poorly studied objects. Also fundamental in our variability studies is VVV's temporal coverage, with $\sim100$ epochs over a 5-year period; the conditions of the observations, with most of them taken with seeing under 1~arcsec; the excellent spatial resolution of the VIRCAM camera, $0\farcs34$ per pixel; and a complete spatial coverage of the inner regions of the Galaxy. VVV data thus place us in a privileged position, as far as the study of variable stars in heavily obscured star clusters goes, allowing us to probe deep into the centers of these objects and out to their tidal radii~-- and beyond. 

\begin{figure*}
\centering
\includegraphics[width=0.8\textwidth]{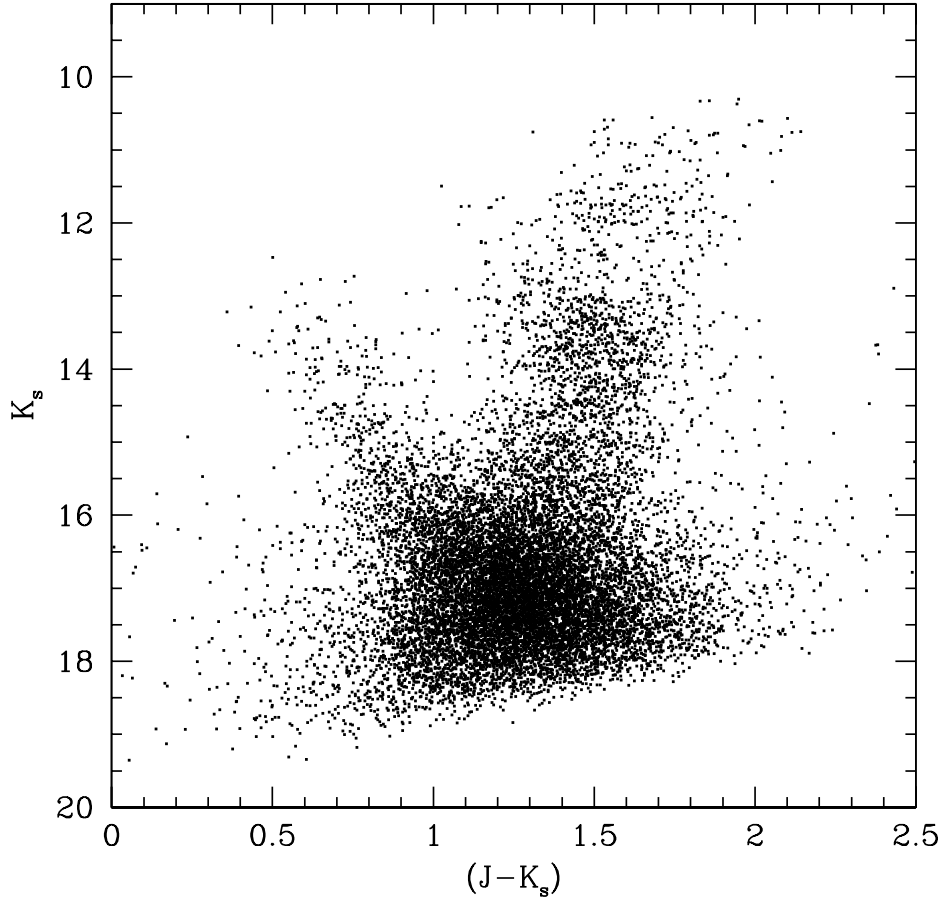}
\vskip0pt
\caption{Near-IR VVV CMD of the inner $1\farcm75$ region of the GC Terzan~10. Note that in the same region of the CMD cluster and bulge field red-giant branch stars co-exist, and that at fainter magnitudes the cluster stars are also mixed with disk main-sequence stars. The GC branches also show some broadening, due to the presence of differential foreground extinction.}
\label{fig_cmd_ter}
\end{figure*}

\begin{figure*}
\centering
\includegraphics[width=0.8\textwidth]{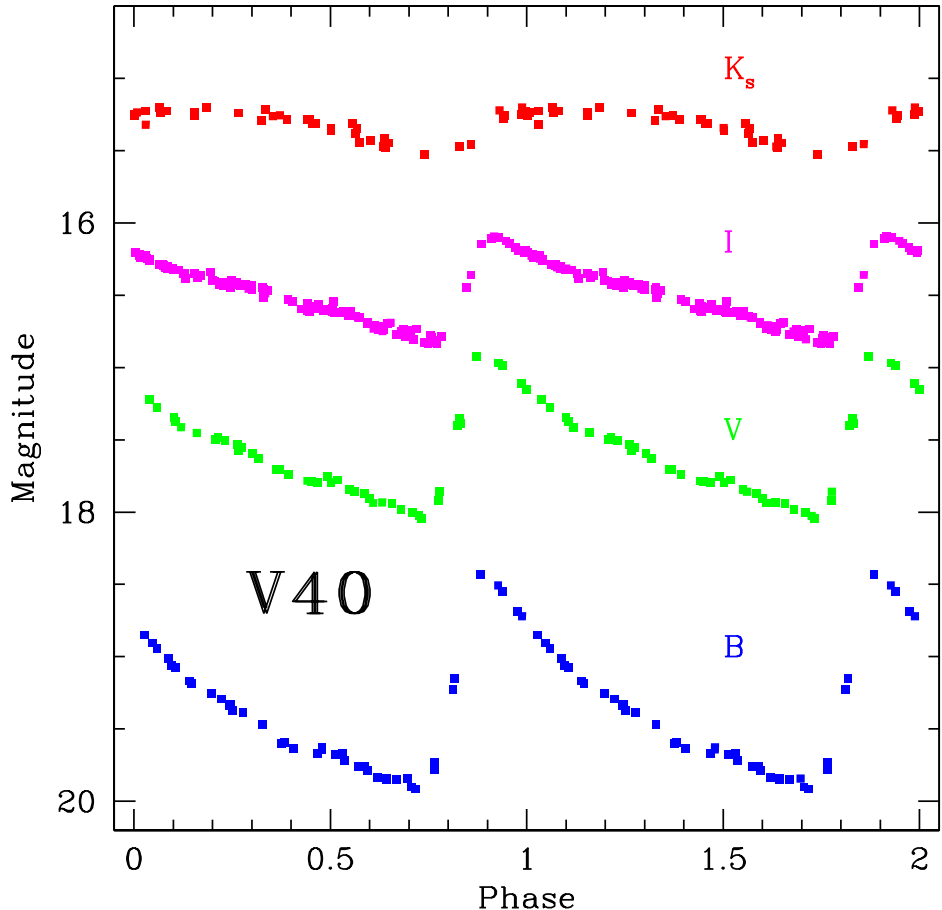}
\vskip0pt
\caption{Light curves at different wavelengths for V40 (an ab-type RR Lyrae star), one of the
  previously known variable stars in the GC NGC~6441. The $K_s$-band 
  light curve comes from the VVV data, while the $I$-band light curve comes
  from the OGLE survey \citep{so11}, and the $B$ and $V$ data from the
  study by \citet{pr01}. We have added 1~mag to the $B$ values
  to separate them more clearly from $V$. Note the change in amplitude
  and shape of the light curve at different wavelengths.}
\label{fig_var_6441}
\end{figure*}

We have already started the search for variable stars in OCs and GCs alike, focusing initially on the VVV fields that lie in the bulge region of the survey, since the latter have been observed more often than the disk fields (\S\ref{sec:var-over}). In what follows, we describe some of our initial results, as far as the GC variability search is concerned \citep[see also][]{al13}.

GCs contain large numbers of RR Lyrae stars, which show a tight relation between periods and absolute magnitude in the near-IR \citep{lo86,cas04,mcea04}, thus helping constrain the VVV GCs distances and extinctions. For many of these clusters, especially the faintest ones, the other photometric technique extensively used to derive their parameters, the study of their color-magnitude diagrams (CMD), is highly complicated by the presence of elevated field stellar contamination and differential reddening, even at near-IR wavelengths (see Fig.~\ref{fig_cmd_ter}). When present, RR Lyrae variables can provide us with better means to obtain these physical parameters. Unfortunately, RR Lyrae analysis in the near-IR has its own difficulties: the amplitude of the RR Lyrae decreases towards the near-IR when compared to the optical, and the shape of the RRab light curves becomes more sinusoidal (see Fig.~\ref{fig_var_6441}). Even so, the quality of our light curves will allow us to look for these variables and to properly characterize their amplitudes, periods, mean magnitudes, and colors. 

In \citet{al13}, we present our first attempts to tackle the study of variable stars in the VVV GCs, and show our efforts to characterize, in the near-IR, variables that had already been previously identified in optical studies (as in the case of NGC~6441), or to discover new variables in extremely obscured GCs where optical observations are very complicated or unfeasible (as in the cases of Terzan~10 or 2MASS-GC02). 

An intriguing characteristic that only Galactic GCs seem to show is the so-called Oosterhoff dichotomy \citep[e.g.,][and references therein]{ca09,sm11}, i.e., Galactic GCs divide themselves in two main groups according to the mean period of their RRab variables, Oosterhoff I systems having shorter periods ($\langle P_{\rm ab}\rangle\sim0.55$~days) and Oosterhoff II systems with longer periods ($\langle P_{\rm ab}\rangle\sim0.64$~days). In the Milky Way, but not in nearby satellite galaxies, very few systems are found in the so-called ``Oosterhoff gap'' zone, with $0.58 \leq \langle P_{\rm ab}({\rm d})\rangle \leq 0.62$. Interestingly, the periods of the RR Lyrae candidates we found in 2MASS-GC02 and in Terzan~10 make these two GCs outliers in the established picture, with Terzan~10 being an Oosterhoff II GC but seemingly having too high a metallicity (${\rm [Fe/H]} = -1.0$~dex; \citealt{ha96}, Feb. 2010 update) to belong to the group, and 2MASS-GC02 falling in the almost empty Oosterhoff gap region \citep{al13}. Terzan~10 could thus be a less extreme example of the new ``Oosterhoff III'' group proposed by \citeauthor{bpea00} \citep[\citeyear{bpea00}; see also][]{pr01,pr02,pr03,tmcea06}, which so far contains exclusively bulge GCs with even higher [Fe/H] and longer $\langle P_{\rm ab}\rangle$, namely NGC~6388 and NGC~6441. Any new RR Lyrae stars that we may be able to find in the inner GCs, and especially those with very few or no RR Lyrae known, will be particularly useful to further explore and understand the Oosterhoff dichotomy, and thus help us place bulge GCs in the wider picture of Milky Way formation \citep[][and references therein]{ca09}.

\begin{figure*}
\centering
\includegraphics[width=0.75\textwidth]{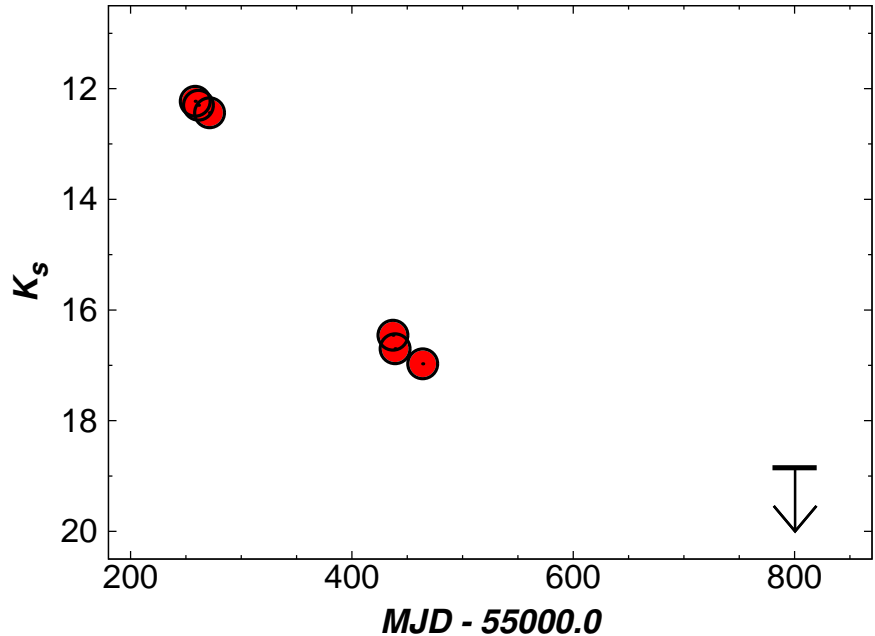}
\vskip0pt
\caption{VVV $K_s$-band light curve of source VVV-WIT-01 \citep{dmea12}, an extreme transient whose classification remains unclear.}
\label{fig:wit01}
\end{figure*}

\subsection{Miscellaneous Variables}\label{sec:misc} 
In addition to the many different types of well-classified variable stars discussed in the previous sections, the VVV Survey data have already resulted in the detection of various transient objects \citep{rsea12a,rsea13}, some of which appear to defy commonly adopted classification schemes. One example is provided by VVV-WIT-01 \citep{dmea12}, an extreme transient event whose light curve is shown in Figure~\ref{fig:wit01}. Its color, as shown by the VVV multi-color data, is also extremely red, with $(J-K_s) > 5$~mag. As discussed by \citeauthor{dmea12}, the source could be an eruptive pre-main sequence star, a reddened luminous blue variable, a nova, or even a highly obscured supernova event or a previously unknown kind of variable star. Like VVV-WIT-01, many additional transient events are expected to be detected in the VVV data, thus opening an exciting path towards the discovery and analysis of energetic events towards the innermost and most heavily obscured regions of the Milky Way.   

\begin{figure*}
\centering
\includegraphics[width=0.925\textwidth]{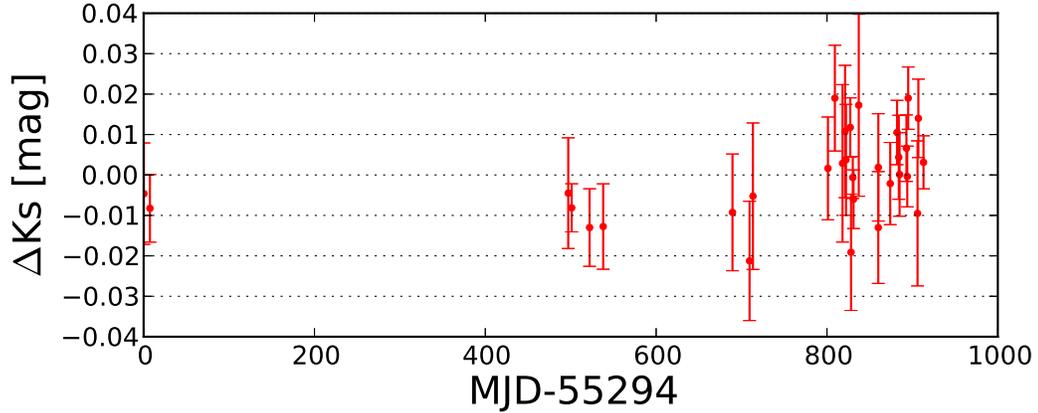}
\vskip0pt
\caption{VVV $K_s$-band differential light curve of VVV-BD-001. Magnitudes are measured with respect to 4 nearby comparison stars with similar mean magnitudes ($\left| K_s^{\rm BD}- K_s^{\rm comp} \right| \leq 0.3$~mag). The light curve was shifted by $+0.015$~mag, so that $\langle \Delta K_s \rangle = 0$~mag. No evidence of periodic variability could be detected in the data \citep[see][]{jcbea13}.}
\label{fig:bd}
\end{figure*}

At the other extreme, long-term photometric monitoring of ultra-cool dwarf (UCD) stars can also be carried out with the VVV Survey. Indeed, the variability of UCDs is key to understand the atmospheric conditions and cloud formation in sub-stellar objects. Several groups have monitored UCDs on a daily basis, over timescales ranging from a few days to a few months \citep[e.g.,][]{daea13,ahea13}. Using VVV data we will be able not only to discover new UCDs, but also probe a new regime of variability for sub-stellar objects, thus helping to constrain long-term atmospheric changes in these very low-mass objects. In Figure~\ref{fig:bd} is given an example of the long-term behavior of the newly discovered brown dwarf \object{VVV-BD-001} \citep{jcbea13}. As discussed by \citeauthor{jcbea13}, the available data do not present signs of periodicity, and any variability is restricted to amplitudes $\Delta K_{\rm s} \leq 0.05$~mag.

\section{Conclusions}\label{sec:concl}
The VVV ESO Public Survey provides a treasure trove of scientific data that can be exploited in numerous different scientific contexts. In terms of stellar variability, the project will provide up to several million calibrated $K_s$-band light curves for genuinely variable sources, including pulsating stars, eclipsing systems, rotating variables, cataclysmic stars, microlenses, planetary transits, and even transient events of unknown nature. At the present point in time, with the data-gathering phase of the VVV Survey having just crossed its half-way mark, we are really just taking the first steps in what will certainly be a long and exciting journey, during which it will be possible to address a myriad of time-domain astronomical applications, including not only research on variable stars as such but also their use as distance indicators and tracers of Galactic structure, origin, and evolution. VVV is a Public Survey, and so the data will quickly be made available to the entire astronomical community as we move along, thus opening the door to many additional applications and synergies with other ongoing and future projects that target the same fields as those covered by VVV.


\vskip 0.5cm

\noindent {\bf Acknowledgements.} The VVV Survey is supported by the European Southern Observatory, 
the Basal Center for Astrophysics and Associated Technologies (PFB-06), and the Chilean Ministry 
for the Economy, Development, and Tourism's Programa Iniciativa Cient\'{i}fica Milenio through 
grant P07-021-F, awarded to The Milky Way Millennium Nucleus. 
We gratefully acknowledge the support provided by Fondecyt through grants \#1110326 (M.C., I.D., 
J.A.-G.), 1120601 (J.B.), 1130140 (R.K.), 3130320 (R.C.R), and 3130552 (J.A.-G.). 
C. Navarrete acknowledges grant CONICYT-PCHA/Mag{\'i}ster Nacional/2012-22121934.  M.C., A.J., S.E., and K.P. acknowledge additional support by project VRI-PUC 25/2011.

\end{document}